\documentclass[aps,pra,twocolumn,superscriptaddress]{revtex4-2}  

\usepackage[dvipsnames]{xcolor}

\definecolor{darkred}{rgb}{0.6,0.05,0.05}
\definecolor{darkgreen}{rgb}{0.05,0.6,0.05}
\definecolor{darkblue}{rgb}{0.05,0.05,0.6}
\usepackage[colorlinks]{hyperref}
\hypersetup{colorlinks,
linkcolor=darkred,
filecolor=darkgreen,
urlcolor=darkblue,
citecolor=darkgreen}

\usepackage{mathtools}
\usepackage{amsmath}
\allowdisplaybreaks
\usepackage{bbold}
\usepackage{amssymb}
\usepackage[normalem]{ulem}
\usepackage{amsthm}
\usepackage{xfrac}
\usepackage{xcolor}
\usepackage{dsfont}
\usepackage{siunitx}
\usepackage{physics}
\usepackage{graphicx}
\usepackage{hyperref}
\usepackage[capitalise]{cleveref}
\usepackage{dcolumn}
\usepackage{bm}
\usepackage{comment}
\usepackage{makecell}
\usepackage{xcolor}

\newcommand{\fil}[1]{{\leavevmode\color{blue}{#1}}}

\newcommand{\rme}{{\rm e}}
\newcommand{\rmd}{{\rm d}}
\newcommand{\rmi}{{\rm i}}

\begin{document}

\author{Leo Kruglikov}
\altaffiliation{These authors contributed equally to this work}
\affiliation{Institute of Physics, \'{E}cole Polytechnique F\'{e}d\'{e}rale de Lausanne (EPFL), CH-1015 Lausanne, Switzerland}
\affiliation{Center for Quantum Science and Engineering, \\ \'{E}cole Polytechnique F\'{e}d\'{e}rale de Lausanne (EPFL), CH-1015 Lausanne, Switzerland}
\author{Filippo Ferrari}
\altaffiliation{These authors contributed equally to this work}
\affiliation{Institute of Physics, \'{E}cole Polytechnique F\'{e}d\'{e}rale de Lausanne (EPFL), CH-1015 Lausanne, Switzerland}
\affiliation{Center for Quantum Science and Engineering, \\ \'{E}cole Polytechnique F\'{e}d\'{e}rale de Lausanne (EPFL), CH-1015 Lausanne, Switzerland}
\author{Vincenzo Savona}
\altaffiliation{Correspondence should be addressed to \fil{filippo.ferrari@epfl.ch} or \fil{vincenzo.savona@epfl.ch}}
\affiliation{Institute of Physics, \'{E}cole Polytechnique F\'{e}d\'{e}rale de Lausanne (EPFL), CH-1015 Lausanne, Switzerland}
\affiliation{Center for Quantum Science and Engineering, \\ \'{E}cole Polytechnique F\'{e}d\'{e}rale de Lausanne (EPFL), CH-1015 Lausanne, Switzerland}

\title{Chaos and quantum regimes in $n$-photon driven, dissipative bosonic chains}

\date{\today}
             
\begin{abstract}
We investigate the steady-state dynamical regimes of boundary-driven, dissipative bosonic chains subjected to $n$-photon drives. Using the truncated Wigner approximation, we explore how multi-photon drives shape the interplay between quantum fluctuations, nonlinear interactions, and dissipative processes in such quantum systems. We identify two main regimes: a chaotic hydrodynamic regime characterized by the restoration of a local $\mathbb{U}(1)$ symmetry, photon saturation due to Kerr nonlinearity, and spatial prethermalization effects; and a non-chaotic resonant nonlinear wave (RNW) regime exhibiting sub-Poissonian photon statistics, persistent $\mathbb{Z}_n$ symmetry, and quantum-driven phase decoherence. Our findings reveal the universal nature of the hydrodynamic regime and highlight the RNW regime’s sensitivity to boundary driving conditions, suggesting novel routes for quantum state engineering in driven-dissipative quantum devices. These results are experimentally relevant for state-of-the-art circuit quantum electrodynamics platforms.
\end{abstract}

\maketitle

\section{Introduction}

Over the past two decades, quantum technologies have evolved from the control of isolated few-level systems to the engineering of complex many-body devices with tailored interactions and environments.
Superconducting circuit platforms \cite{krantz_quantum_2019, blais_circuit_2021} in particular allow the construction of extended networks of nonlinear resonators with exquisite control over drive and dissipation, opening a new frontier for nonequilibrium quantum matter. This setting enabled the experimental investigation of many-body phases of matter \cite{satzinger_realizing_2021, zhang_superconducting_2023, rosenberg_dynamics_2024, andersen_thermalization_2025}, the coherent manipulation of atom-photon bound states \cite{liu_quantum_2017, sundaresan_interacting_2019, scigliuzzo_controlling_2022}, the study of quantum chaotic dynamics \cite{braumuller_probing_2022}, as well as the implementation of quantum error correction schemes \cite{ofek_extending_2016, krinner_realizing_2022, google_quantum_ai_suppressing_2023, google_quantum_ai_quantum_2025, putterman_hardware-efficient_2025}.

The interaction with the surrounding environment is unavoidable when quantum devices are subject to driving mechanisms or readout operations.
Far from being only a detrimental effect for quantum technologies, dissipation can steer the quantum dynamics stabilizing certain types of superconducting qubits, like dissipative cat qubits \cite{mirrahimi_dynamically_2014, leghtas_confining_2015, touzard_coherent_2018, lescanne_exponential_2020, reglade_quantum_2024, rousseau_enhancing_2025}, or inducing phenomena that are different or simply absent in the isolated scenario, such as dissipative criticality \cite{carusotto_quantum_2013, carmichael_breakdown_2015, bartolo_exact_2016, fink_observation_2017, biondi_nonequilibrium_2017, fink_signatures_2018, minganti_spectral_2018, lieu_symmetry_2020, gravina_critical_2023} and dissipative quantum chaos \cite{akemann_universal_2019, hamazaki_universality_2020, sa_complex_2020, prasad_dissipative_2022, dahan_classical_2022, 
kawabata_symmetry_2023,
villasenor_breakdown_2024,
ferrari_dissipative_2025}.
Specifically, a rich dynamical landscape generated by the competition among Hermitian time evolution, driving and dissipation mechanisms appears already in a minimal one dimensional chain of coupled nonlinear oscillators with boundary drive and dissipation.
This setup has been realized in circuit QED architectures \cite{FitzpatrickPRX17, FedorovPRL21} and investigated theoretically in the classical limit \cite{debnath_nonequilibrium_2017, prem_dynamics_2023, kumar_observation_2024} and including quantum fluctuations \cite{ferrari_chaos_2024}.
In particular, Ref.~\cite{ferrari_chaos_2024} identified two notable dynamical regimes in the nonequilibrium steady state (NESS). 
The first one is a chaotic regime where the local density matrices are effectively described by equilibrium Gibbs states.
Along the chain, a two-stage thermalization mechanism leaves room for an extended ``prethermal" domain characterized by photon saturation and anomalous heating, before the consequent thermalization at the end of the chain, where the local states are vacuum-like states dressed by thermal fluctuations.
The second regime, described classically in Refs.~\cite{prem_dynamics_2023, kumar_observation_2024} and quantum mechanically in Ref.~\cite{ferrari_chaos_2024}, is the resonant nonlinear wave (RNW) regime.
Here, quantum fluctuations are relevant to shape the local states, leading to sub-Poissonian statistics of the bosonic field and phase decoherence absent in the classical limit. 

In this work, we address how boundary driving shapes the bulk NESS properties of chains comprising tens to hundreds of nonlinear oscillators, using the truncated Wigner approximation (TWA) \cite{carmichael_statistical_1999, polkovnikov_phase_2010}.
This is a relevant question for quantum state engineering \cite{diehl_quantum_2008, harrington_engineered_2022, landi_nonequilibrium_2022}, where simple local operations like boundary drives can determine the global structure of correlated steady states.
We explore this systematically by applying a coherent $n$-photon drive at the boundary, which imposes a weak $\mathbb{Z}_n$ symmetry in the system.
Such driving mechanisms are both technologically relevant and conceptually rich--offering connections to spontaneous symmetry breaking and quantum criticality \cite{rota_quantum_2019, verstraelen_gaussian_2020,minganti_dissipative_2023, beaulieu_observation_2025}, and enabling applications in quantum sensing \cite{di_candia_critical_2023, alushi_collective_2025, beaulieu_criticality-enhanced_2025}, quantum memories \cite{labay-mora_quantum_2023, labay-mora_quantum_2024}, and cat-qubit control \cite{puri_engineering_2017, puri_stabilized_2019, puri_bias-preserving_2020, grimm_stabilization_2020, frattini_observation_2024}.

\begin{figure*}[t!]
\centering
\includegraphics[width=0.8 \textwidth]{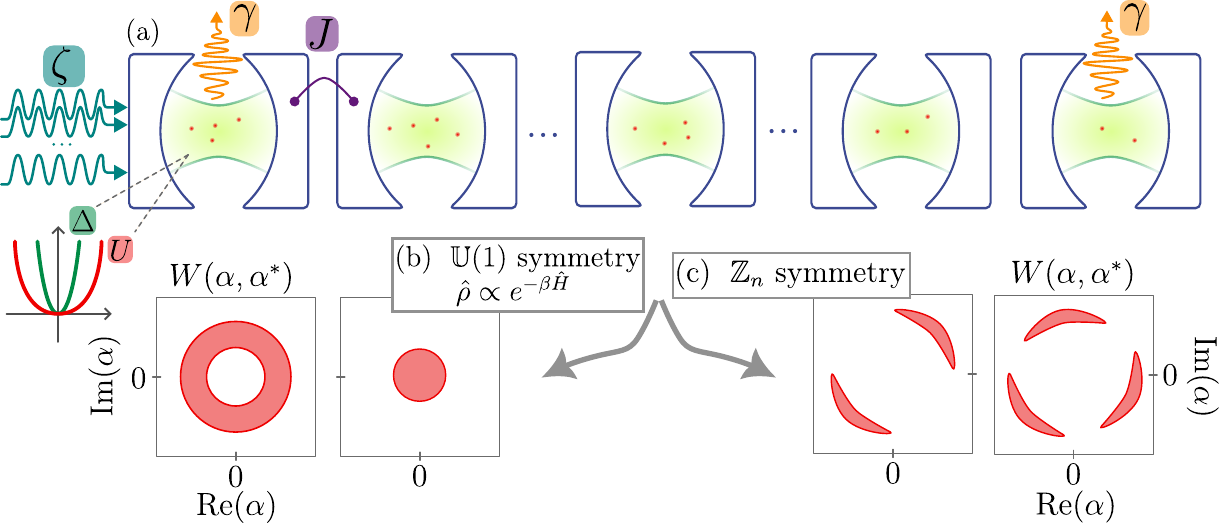}\vspace{0.8em}
\caption{
Sketch of the system and summary of the main results. (a) Chain of $L$ coupled nonlinear resonators subject to boundary single-photon losses and to a $n$-photon drive at the leftmost site, giving to the system Hamiltonian a global $\mathbb{Z}_n$ symmetry.
(b) In the presence of chaos, the NESS reaches thermal equilibrium in the spatial direction, where the $\mathbb{Z}_n$ symmetry is replaced by a local $\mathbb{U}(1)$ symmetry. 
The local quantum states are described by a Gibbs state with effective local temperature and chemical potential. 
A transition between a prethermal domain characterized by annular local Wigner function, large, positive chemical potential and anomalous heating to a thermal domain with a bell-shaped local Wigner function, negative chemical potential and decreasing temperature is observed.
(c) In the absence of chaos, the system reaches a far-from-equilibrium steady state dubbed resonant nonlinear wave regime (RNW) where the symmetry of the boundary drive shapes the local quantum states throughout the chain.
In the RNW regime, quantum fluctuations determines the fundamental features of the local states.
}
\label{fig:artistic_picture}
\end{figure*}

We find that the sensitivity of the bulk state to the boundary drive depends strongly on the underlying dynamical regime.
In the presence of chaos, long-wavelength hydrodynamic modes rapidly erase any trace of the imposed $\mathbb{Z}_n$ symmetry, which remains localized near the driven edge.
This indicates that the emergence of spatial prethermalization and local thermal equilibrium is a robust feature of boundary-driven 1D bosonic systems, insensitive to the symmetry or structure of the drive.
By contrast, in the RNW regime, the entire chain inherits the boundary-imposed $\mathbb{Z}_n$ symmetry.
Local states along the chain develop $n$-modal quasi-probability distributions in phase space, exhibiting quantum-induced phase decoherence and radial localization.
This regime provides a clear example of a correlated NESS whose global features are dictated by the boundary—offering a promising route for the design of novel quantum phases in extended driven-dissipative systems.
Our predictions can be verified in state-of-the-art driven-dissipative circuit QED devices.

The paper is organized as follows. 
In Sec.~\ref{sec:model} we discuss the physical model and the methods considered in this work.
In Sec.~\ref{sec:2_photon} we study the 2-photon driven bosonic chain.
In Sec.~\ref{sec:3_photon} we study the 3-photon driven bosonic chain.
Sec.~\ref{sec:conclusion} presents the conclusions and outlook.
In the Appendices, we discuss in detail the main methods, namely the TWA, the semiclassical OTOCs and the fitting procedures used to assess hydrodynamic behavior.

\section{Model and methods}\label{sec:model}

\subsection{$n$-photon driven, dissipative Bose-Hubbard chain}

We consider the one-dimensional lattice of nonlinear coupled resonators sketched in Fig.~\ref{fig:artistic_picture}.
We assume that the leftmost resonator is driven by a continuous $n$-photon drive.
In the frame rotating at the frequency of the driving field, the Hamiltonian governing the system is a driven Bose-Hubbard chain, written in terms of $L$ annihilation (creation) bosonic many-body operators $\hat{a}_\ell$ ($\hat{a}_\ell^\dagger$),
\begin{align}\label{eqs:hamiltonian}
    \hat{H} =& \sum_{\ell=1}^L\left(-\Delta \, \hat{a}_\ell^{\dagger}\hat{a}_\ell + \frac{1}{2}\, U\,\hat{a}_\ell^{\dagger}\hat{a}_\ell^{\dagger}\hat{a}_\ell\hat{a}_\ell\right)\\* &- J\sum_{\ell=1}^{L-1}\big{(}\hat{a}_{\ell+1}^{\dagger}\hat{a}_\ell + \hat{a}_\ell^{\dagger}\hat{a}_{\ell+1}\big{)} + \frac{\zeta}{n}(\hat{a}_1^{\dagger n} + \hat{a}_1^n)\,, \nonumber
\end{align}
where $\Delta = \omega_\textrm{d}-\omega_0$ represents the frequency detuning between the drive and cavity resonance, $U$ quantifies the Kerr nonlinear strength, $J$ is the nearest-neighbor hopping coefficient and $\zeta$ is the amplitude of the $n$-photon driving field.
We assume that the system experiences sizable single-photon losses only at the edges of the array, while the bulk remains dissipationless.
The dynamics of the many-body density matrix $\hat{\rho}$ is governed, in the Markov approximation, by the Lindblad master equation \cite{breuer_theory_2007}
\begin{equation}\label{eqs:lindblad}
    \frac{\partial\hat{\rho}}{\partial t} = -\rmi[\hat{H}, \hat{\rho}] +
       \mathcal{D}[\hat L_1] \hat \rho +    \mathcal{D}[\hat L_L]  \hat \rho\,,
\end{equation}
where 
\begin{equation}\label{eqs:dissipator}
    \mathcal{D}[\hat L_\ell] \hat \rho := \hat L_\ell \hat{\rho} \hat L_\ell^\dagger - \frac{1}{2}\left\{\hat L_\ell^{\dagger}\hat L_\ell, \hat\rho \right\},
\end{equation}
is the Lindblad dissipator and $\hat{L}_\ell = \sqrt{\gamma}\hat{a}_\ell$ are the jump operators describing single-photon loss mechanisms, and $\gamma$ is the single-photon loss rate.
Given the absence of strong Liouvillian symmetries in Eq.~\eqref{eqs:lindblad} for the system here considered, the master equation~\eqref{eqs:lindblad} admits a unique nonequilibrium steady state (NESS), $\hat{\rho}_{\rm ss} = \lim_{t\to\infty}\hat{\rho}(t)$, which is independent on the chosen initial conditions \cite{breuer_theory_2007}.
Throughout the paper, we set $\gamma$ as the unit of energy and we consider the weakly nonlinear regime by imposing $U=0.1$. 

The proposed model can be implemented in state-of-the-art circuit QED devices, where weak photon-photon interactions are engineered via Josephson junction arrays embedded in superconducting resonators \cite{FitzpatrickPRX17, FedorovPRL21}. Moreover, multi-photon drive terms can be realized through nonlinear elements such as superconducting quantum interference devices, which enable tunable nonlinearities and parametric couplings \cite{wilson_photon_2010, krantz_investigation_2013}.

\subsection{Truncated Wigner approximation}\label{sec:twa}

We simulate the many‑body dynamics governed by the Lindblad master equation Eq.~\eqref{eqs:lindblad} using the truncated Wigner approximation (TWA). In this approach, the density matrix is represented by its Wigner quasiprobability distribution, which in general satisfies an infinite‑order partial differential equation. The TWA truncates this expansion at second order in derivatives—equivalently, at first order in $\hbar$ beyond the classical limit—thereby reducing the evolution to a Fokker–Planck equation. Finally, one converts that Fokker–Planck equation into a set of stochastic (Langevin) equations for the complex fields $\alpha_\ell$. Physically, this approximation incorporates leading-order quantum fluctuations on top of the classical (mean‑field) dynamics \cite{carmichael_statistical_1999, polkovnikov_phase_2010}.
The coupled stochastic differential equations read
\begin{align}\label{eqs:stochastic_differential_equations}
    \rmi \frac{\partial  \alpha_1}{\partial t} &=   -f(\alpha_1)  - J  \alpha_{2} + F(\alpha_1^*)^{n-1} - \frac{\rmi\gamma}{2}\alpha_1 + \sqrt{\frac{\gamma}{2}} \xi_1(t)
    \,, \nonumber \\
    \rmi \frac{\partial  \alpha_\ell}{\partial t} &= -f(\alpha_\ell)  - J (\alpha_{\ell-1} + \alpha_{\ell+1})
    \,,\ l=2,...,L-1 \nonumber  \\
    \rmi \frac{\partial  \alpha_L}{\partial t} &=  -f(\alpha_L)  - J  \alpha_{L-1}  - \frac{\rmi\gamma}{2}\alpha_L + \sqrt{\frac{\gamma}{2}} \xi_L(t)
    \,,
\end{align}
where $f(\alpha) = \Delta \, \alpha - U\,(|\alpha|^2-1)\alpha$, $\xi_1$ and $\xi_L$ are complex-valued Gauss-correlated white-noise random processes such that $\langle\xi_1(t)\rangle =\langle \xi_L(t)\rangle=0$ and $\langle\xi_1(t)\xi_1^*(t')\rangle = \langle\xi_L(t)\xi_L^*(t')\rangle =\delta(t-t')$. 
The TWA is particularly suited for weakly nonlinear bosonic systems close to the classical limit \cite{carmichael_statistical_1999, polkovnikov_phase_2010}.
In the Appendix \ref{sec:appendix_A1} we discuss the derivation of the TWA equations starting from the phase-space formalism of quantum mechanics.
Within this formalism, expectation values of operators can be mapped onto statistical expectation values averaged over many solutions of Eqs.~\eqref{eqs:stochastic_differential_equations}.
We refer the Reader to the Appendix \ref{sec:appendix_A1} for a more in depth discussion.
In the analysis that follows, unless otherwise stated, we will always assume the initial state to be the bosonic vacuum $\hat{\rho}(0) = \bigotimes_{\ell=1}^L\ketbra{0}_\ell$.

The TWA has been employed for the successful characterization of dissipative phase transitions, dissipative time crystals, chaotic dynamics and transport phenomena, in driven-dissipative bosonic lattices \cite{dagvadorj_nonequilibrium_2015, dujardin_elastic_2015, dujardin_breakdown_2016, vicentini_critical_2018, vicentini_optimal_2019, deuar_fully_2021, ferrari_chaos_2024}. 
In Appendix \ref{sec:appendix_A2} we validate the use of the TWA by comparing it with full quantum dynamics on small system sizes.
We also discuss the comparison between the TWA and the classical limit of Eq.~\eqref{eqs:lindblad}, coinciding with the Gross-Pitaevskii equations of motion, where all quantum fluctuations are neglected (see Appendix \ref{sec:appendix_A4}).

\subsection{Semiclassical OTOC}

In classical mechanics, the distinction between regular and chaotic dynamics is based on the concept of Lyapunov instability \cite{strogatz_nonlinear_2018}: two nearby trajectories at the early stage of the dynamics exponentially deviate in time at a rate governed by the largest Lyapunov exponent $\lambda$.
An early exponential growth characterizes also classical and quantum four-point functions known as out-of-time order correlators (OTOCs).
OTOCs are routinely used tools in classical and quantum many-body systems to diagnose and characterize the scrambling of quantum and classical information \cite{maldacena_bound_2016, swingle_measuring_2016, hashimoto_out--time-order_2017, bohrdt_scrambling_2017, das_light-cone_2018, bilitewski_temperature2018, manas_spatiotemporal_2020, bilitewski_classical2021, deger_arresting2022, xu_scrambling_2024}.
In this study, we consider the semiclassical version of the OTOC between the number and phase degrees of freedom as a probe of chaotic dynamics. 
Following Ref.~\cite{ferrari_chaos_2024} we write the OTOC as
\begin{equation}\label{eqs:semiclassical_OTOC}
    D_{k,\ell}(\tau) := 1 - \lim_{t\to\infty}\left\langle \cos\left[ \varphi_\ell^{(a)}(t+\tau) - \varphi_\ell^{(b)}(t+\tau)\right]\right\rangle,
\end{equation}
where $\varphi_\ell(t) = \textrm{arg}[\alpha_\ell(t)]$ is the phase computed along single trajectories. 
The superscripts $(a)$ and $(b)$ denote two replicas of the system that are equal until time $t$.
At time $t$, we apply a perturbation $|\varepsilon|\ll1$ to the phase of $\alpha_k(t)$ of replica $b$, \textit{i.e.}, $\varphi_k^{(b)}\to\varphi_k^{(b)}+\varepsilon$.
Replica $\varphi^{(b)}_\ell$ therefore depends implicitly on the $k$ index.
The time evolution of the two replicas is then computed with the same noise realization and the average in Eq.~\eqref{eqs:semiclassical_OTOC} is computed over many instances of the noise.
Throughout the paper, we fix $\varepsilon=10^{-2}$.

If the dynamics is regular, $\varphi_\ell^{(a)}$ and $\varphi_\ell^{(b)}$ are expected to remain correlated over time, $\varphi_\ell^{(a)}\simeq\varphi_\ell^{(b)}$, leading to $D_{k,\ell}( \tau \to \infty) \simeq 0$.
If the dynamics is instead chaotic, a rapid decorrelation of $\varphi_\ell^{(a)}$ and $\varphi_\ell^{(b)}$ is expected, and the average is taken over uniform random numbers in the interval $[-1,\,1]$, leading to $D_{k,\ell}(\tau \to \infty) \simeq 1$.
The exponential growth of $D_{k,\ell}(\tau)$ at initial times $\tau$ can be considered a defining feature of semiclassical chaos in the steady state of the driven-dissipative Bose-Hubbard chain.
In Appendix \ref{sec:appendix_A3} we thoroughly characterize the OTOC dynamics.

\section{2-photon driven bosonic chain}\label{sec:2_photon}

\begin{figure*}[t!]
\centering
\includegraphics[width=0.95 \textwidth]{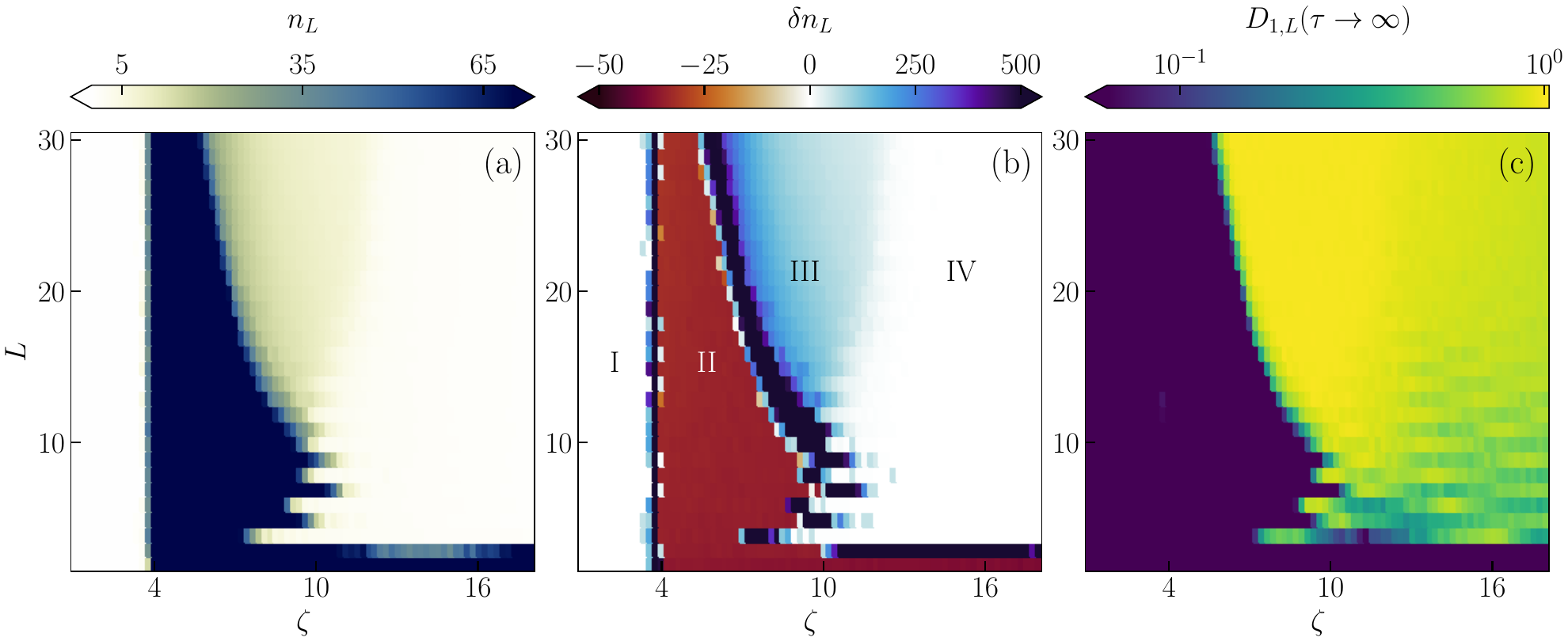}\vspace{0.8em}
\caption{
Phase diagram in the NESS as a function of the chain length $L$ and the 2-photon drive amplitude $\zeta$ for the last resonator $\ell=L$.
(a) Photon number $n_L$.
(b) Photon fluctuations $\delta n_L$, computed according to Eq.~\eqref{eqs:photon_fluctuations}.
(c) Saturation value of the semiclassical OTOC $D_{1,L}(\tau\to\infty)$, computed according to Eq.~\eqref{eqs:semiclassical_OTOC}.
The four regimes labeled I, II, III, and IV, identified in panel (b) are discussed in the text.
Results for panels (a) and (b) are computed by averaging over $N_{\rm traj} = 5\times 10^2$ independent trajectories, while results in panel (c) over $10^2$ trajectories. 
Statistics are further improved by averaging over a time window $\Delta \tau = 10^3$ after reaching the steady state.
Throughout the manuscript, we fix $\gamma=1$ and $U=0.1$. In this section, we instead fix $\Delta=5.6$ and $J=2.2$.
}
\label{fig:phase_diagram}
\end{figure*}

In this section, we consider parametrically, boundary-driven, boundary-dissipative Bose-Hubbard chains, by choosing $n=2$ in Eq.~\eqref{eqs:hamiltonian}. We consider the following parameters as fixed: $\Delta=5.6$, $J=2.2$. We find that the qualitative phase diagram of the system does not change for other choices of these parameters.

\subsection{Phase diagram}\label{sec:phase_diagram}

We explore the NESS phase diagram by varying the chain length $L$ and the 2-photon drive amplitude $\zeta$.
We characterize the different regimes from three main quantities, all computed at the last site $\ell=L$ and in the steady state. First, we analyze the photon number $n_L=\langle\hat{a}^\dagger_L\hat{a}_L\rangle$. Second, we consider the photon fluctuations, that we measure by computing the quantity
\begin{equation}\label{eqs:photon_fluctuations}
    \delta n_L = \langle\hat{a}_L^{\dagger 2}\hat{a}_L^2\rangle - \langle\hat{a}_L^{\dagger}\hat{a}_L\rangle^2.
\end{equation}
$\delta n_L$ can be directly related to the second-order coherence function $g^{(2)}_L$ via $\delta n_\ell = n_\ell^2(g^{(2)}_\ell-1)$ \footnote{with respect to $g^{(2)}_\ell$, $\delta n_L$ is less prone to numerical instability when $n_\ell$ is close to zero, and a larger number of stochastic trajectories can be needed to reach convergence. For large chains, this implies a significant computational overhead.}.
$\delta n_L$ (like $g^{(2)}_\ell$) quantifies the distance from a Poissonian statistics of the bosonic field: $\delta n_L=0$ identifies the Poissonian statistics typical of coherent states, $\delta n_L>0$ is related to a super-Poissonian statistics that can be associated with fluctuating classical fields, and $\delta n_L<0$ describes a sub-Poissonian statistics signaling nonclassical features \cite{walls_quantum_2008}.

The third quantity we consider is the semiclassical OTOC $D_{1,\ell}(\tau\to\infty)$ defined in Eq.~\eqref{eqs:semiclassical_OTOC}.
The results for $n_L$, $\delta n_L$ and $D_{1,\ell}(\tau\to\infty)$ as a function of $L$ and $\zeta$ are reported in Fig.~\ref{fig:phase_diagram}.
Based on the features of the three quantities described above, we are able to distinguish four regimes, that we order according to the increase of $\zeta$:
\begin{enumerate}
    \item[(I)] \textit{Regular quasilinear regime}. At very low values of $\zeta$, $n_L$ and $\delta n_L$ are close to zero, and the OTOC indicates regular dynamics. In this regime, the local states are squeezed vacua and the effect of nonlinearity is negligible.
    \item[(II)] \textit{Resonant nonlinear wave (RNW) regime}. For $\zeta$ larger than a critical value $\zeta^*$, the photonic population transitions from the vacuum-like state to a highly populated state. 
    The photon number statistics becomes sub-Poissonian and the dynamics is regular, as indicated by $\delta n_L<0$ and $D_{1, L}(\tau\to\infty)\simeq0$.
    These are clean signatures of the resonant nonlinear wave regime \cite{ferrari_chaos_2024}.
    \item[(III)] \textit{Hydrodynamic regime}. By increasing the chain length $L$ the system enters in a chaotic phase characterized by large super-Poissonian fluctuations and a saturated steady-state OTOC $D_{1, L}(\tau\to\infty)$.
    \item[(IV)] \textit{Thermal vacuum regime}. At stronger 2-photon drive amplitudes $\zeta$, the occupation of the last site of the system is approximately zero, and the associated fluctuations are nearly Poissonian, accompanied by a large but not saturated value of $D_{1, L}(\tau\to\infty)$. 
\end{enumerate}

\begin{figure*}[t!]
\centering
\includegraphics[width=1 \textwidth]{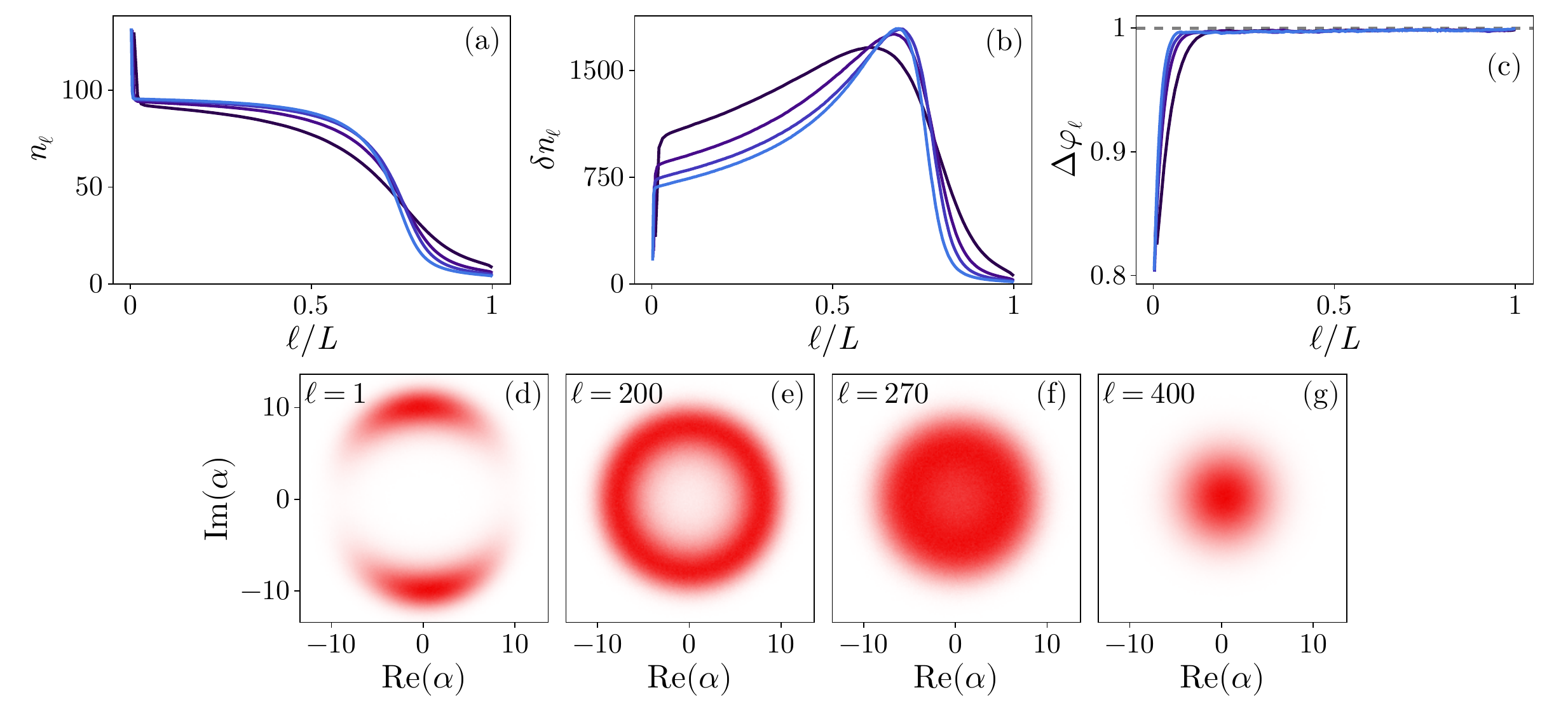}\vspace{0.8em}
\caption{
Two stage relaxation in space.
(a) Photon number $n_\ell$, (b) photon fluctuations $\delta n_\ell$, (c) circular variance $\Delta\varphi_\ell$ for different system sizes $L=100, 200, 300, 400$ (from dark to light blue).
Results are computed by averaging over $N_{\rm traj} = 225$ independent trajectories. 
Statistics are further improved by averaging over a time window $\Delta \tau = 10^4$ after reaching the steady state.
(d-g) Local Wigner functions $W_\ell(\alpha, \alpha^*)$ for the sites (d) $\ell=1$, (e) $\ell=200$, (f) $\ell=283$, and (g) $\ell=400$ for a $L=400$ chain.
The drive amplitude is set to $\zeta=6$.
The other parameters are set as in Fig.~\ref{fig:phase_diagram}.
}
\label{fig:chaos}
\end{figure*}

In the main text, we focus on regimes (II) and (III) as they are characterized by the emergence of interesting quantum features and by the spatial prethermalization identified in Ref.~\cite{ferrari_chaos_2024}.
Regimes (I) and (IV) are less interesting from this point of view, since the local states are either coherent or squeezed states with vanishing population or vacua dressed by classical thermal fluctuations, and they are studied in Appendix \ref{sec:appendix_D}.

With respect to the findings of Ref.~\cite{ferrari_chaos_2024}, where a single-photon drive is considered, our results indicate that a higher-photon drive does not qualitatively change the hydrodynamic regime.
This is expected, since in this regime, only long-wavelength hydrodynamic modes dictate the physics of the system bulk and undriven edge.
We conjecture that the features of the hydrodynamic regime are universal in this class of systems and independent on the details of the boundary driving conditions.
The $\mathbb{Z}_n$ symmetry of the drive, instead, is crucial in shaping the sub-Poissonian RNW regime, compared with the single-photon drive case analyzed in Ref.~\cite{ferrari_chaos_2024}.

\subsection{Hydrodynamic regime}\label{sec:chaotic_regime}

We first focus on the hydrodynamic regime by fixing the 2-photon drive amplitude to $\zeta=6$.
We decompose the bosonic field in the number and phase degrees of freedom, $\alpha_\ell = |\alpha_\ell|\rme^{\rmi\varphi_\ell}$ and we analyze their spatial behavior across driven-dissipative chains of increasing length $L$.

In Fig.~\ref{fig:chaos} we study the photon number $n_\ell$, the photon fluctuations by means of $\delta n_\ell$ and the phase fluctuations by means of the circular variance $\Delta \varphi_\ell^{(1)} = 1-|\langle\rme^{\rmi\varphi_\ell}\rangle|$ in chains with increasing length $L$. 
The circular variance quantifies the phase spreading, and is bounded between 0 (a point in the complex plane) and 1 (uniformly distributed random phases in the $[-\pi, \pi]$ interval) \footnote{We anticipate here that the circular variance can be defined at different orders $\Delta\varphi_\ell^{(m)} = 1 - C_m$, where $m$ is a positive integer and $C_m = |\langle\rme^{\rmi m \varphi_\ell}\rangle|$ is the circular momentum of order $m$. We will introduce these quantities in detail in Sec.~\ref{sec:nrw_regime}. We verified that the circular variance saturates to 1 in the hydrodynamic regime independently of $m$.}.
Results in Figs.~\ref{fig:chaos} (a-c) show a two-stage relaxation process in the spatial direction.
After a microscopic length scale the phase degree of freedom relaxes to a uniform distribution on the circle, as indicated by the saturation of $\Delta\varphi_\ell^{(1)}$.
Within this length scale, the local states inherit the $\mathbb{Z}_2$ symmetry of the driving field, as shown by the local Wigner function $W_\ell(\alpha, \alpha^*)$ for $\ell=1$ reported in Fig.~\ref{fig:chaos} (d), which is manifestly bimodal.
The $\mathbb{U}(1)$ symmetry of the bulk Hamiltonian is subsequently restored over the remaining macroscopic sector of the chain.
This is the fingerprint of a hydrodynamic behavior in the system, as we discuss below.
In Figs.~\ref{fig:chaos} (e-g) we display $W_\ell(\alpha, \alpha^*)$ in the bulk and at the undriven edge of the chain, which consists in a $\mathbb{U}(1)$-symmetric ring or a $\mathbb{U}(1)$-symmetric disk.

\begin{figure*}[t!]
\centering
\includegraphics[width=1 \textwidth]{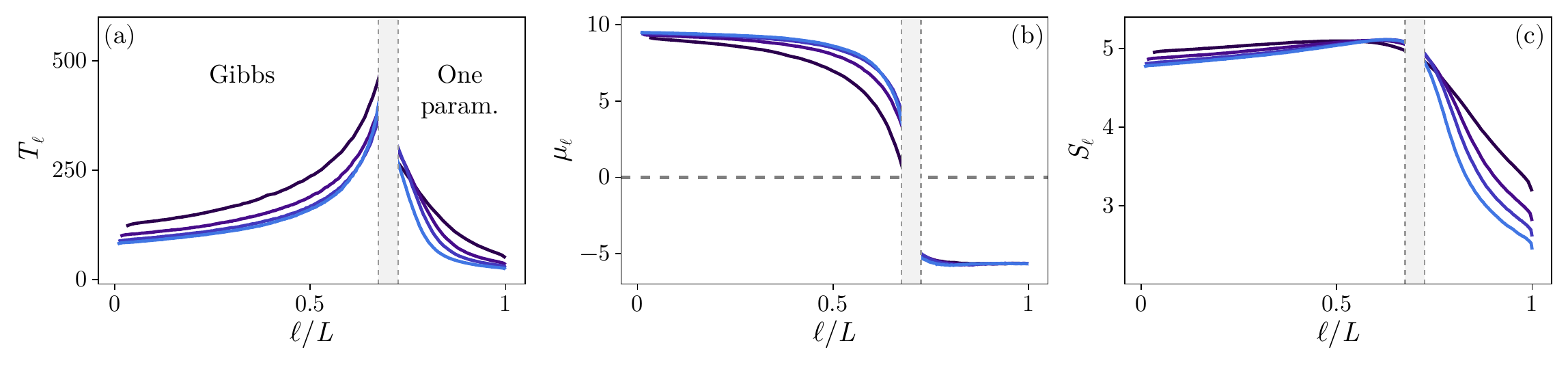}\vspace{0.8em}
\caption{
Hydrodynamic description.
(a) Effective temperature $T_\ell$, (b) effective chemical potential $\mu_\ell$, (c) local entropy $S_\ell$ for different system sizes $L=100, 200, 300, 400$ (from dark to light blue).
The vertical grey dashed line separates the portion of the chain where the quantities are extracted imposing the Gibbs state ansatz \eqref{eqs:gibbs}, or the single-parameter ansatz \eqref{eqs:one_param}.
In this case, the temperature is estimated according to the equipartition theorem as $T_\ell=|\Delta|\langle p_\ell^2\rangle$, being $p_\ell = \sqrt{2}\textrm{Im}(\alpha_\ell)$ the local momentum (see Appendix \ref{sec:appendix_B}).
Results are computed by averaging over $N_{\rm traj} = 225$ independent trajectories. 
Statistics are further improved by averaging over a time window $\Delta \tau = 10^4$ after reaching the steady state.
The drive amplitude is set to $\zeta=6$.
The other parameters are set as in Fig.~\ref{fig:phase_diagram}.
}
\label{fig:hydrodynamics}
\end{figure*}

The number degree of freedom and its fluctuations show instead a slow relaxation, over a macroscopic length scale of the order of the system size.
The photon number $n_\ell$ relaxes to a vacumm-like state only at the end of the chain, remaining approximately constant in the bulk.
The photon fluctuations $\delta n_\ell$ show instead a non-monotonic behavior: they slowly increase away from the boundary drive, and rapidly drop to the Poissonian limit at the end of the chain.
At the point where $\delta n_\ell$ is maximal, $\ell=270$, the ring closes and $W_\ell(\alpha, \alpha^*)$ reaches the maximal extension in phase space.
This change in the local Wigner function topology is connected to different thermodynamic domains within the chaotic chain.

To probe the local equilibria away from the driven boundary, we follow Ref.~\cite{ferrari_chaos_2024} and we inspect whether the local density matrix is given by a Gibbs state reading
\begin{equation}\label{eqs:gibbs}
    \hat{\rho}^{\rm eq}_\ell = \frac{\exp[-(\hat h - \mu_\ell \, \hat a^\dagger \hat a)/T_\ell]}{\operatorname{Tr}\left(\exp[-(\hat h - \mu_\ell \, \hat a^\dagger \hat a)/T_\ell]\right)}\,,
\end{equation}
where $T_\ell$ and $\mu_\ell$ are the effective local temperature and local chemical potential, respectively.
The Hamiltonian $\hat{h} = U(\hat{a}^\dagger\hat{a})^2/2$ represents instead the on-site Kerr nonlinearity of Eq.~\eqref{eqs:hamiltonian}.
In practice, we extract the local thermodynamic quantities by fitting $W_\ell(\alpha, \alpha^*)$ with the Gibbs ansatz \eqref{eqs:gibbs}.
The fitting procedure is described in the Appendix \ref{sec:appendix_A4}.
We find that, as soon as the $\mathbb{U}(1)$ symmetry is restored, the local states throughout the bulk of the chain are effective Gibbs states.
In the undriven tail of the chain, we find that the Wigner function becomes indistinguishable from a thermal Gaussian.
This implies an effective $U=0$ in Eq.~\eqref{eqs:gibbs}, and the Gibbs state reduces to a one-parameter ansatz
\begin{equation}\label{eqs:one_param}
    \hat{\rho}^{\rm eq}_\ell = \frac{\rme^{\xi_\ell\hat{a}^\dagger\hat{a}}}{\operatorname{Tr}\left(\rme^{\xi_\ell\hat{a}^\dagger\hat{a}}\right)},
\end{equation}
with $\xi_\ell=\mu_\ell/T_\ell$ as the only fitting parameter.
To estimate the temperature, given the Gaussian form of the Wigner function of the local states, we resort to the equipartition theorem \cite{toda_statistical_1983} which states that for a 1D system in local thermal equilibrium $T_\ell = |\Delta|\langle p_\ell^2 \rangle$ being $p_\ell = \sqrt{2}\textrm{Im}(\alpha_\ell)$ the local momentum.
We discuss the applicability of the equipartition theorem in Appendix \ref{sec:appendix_B}.
We extract the effective chemical potential from the fitted $\xi_\ell$ and the temperature $T_\ell$ obtained from the equipartition theorem.
Finally, we also monitor the local entropy $S_\ell = -\operatorname{Tr}[\hat{\rho}_\ell\log\hat{\rho_\ell}]$.

In Fig.~\ref{fig:hydrodynamics} (a-c) we plot the spatial profile of $T_\ell$, $\mu_\ell$ and $S_\ell$ for different system sizes.
The shaded area in the three panels separate the region where the Gibbs state with a finite Kerr nonlinearity reproduces the shape of $W_\ell(\alpha, \alpha^*)$, from the region where $U=0$ and the local states are indistinguishable from Gaussian thermal states.
Overall, the behavior of these quentities, as well as the topology of $W_\ell(\alpha, \alpha^*)$, motivate the distinction of three different domains in the driven-dissipative chain:
\begin{enumerate}
    \item [(i)] \textit{$\mathbb{Z}_2$-symmetric nonthermal domain}. In proximity of the coherent drive, the local states inherit the $\mathbb{Z}_2$ symmetry and the local states are not described by Eq.~\eqref{eqs:gibbs}.
    \item [(ii)] \textit{$\mathbb{U}(1)$-symmetric prethermal domain}.
    The Gibbs state \eqref{eqs:gibbs} describes the local physics of the system. The Kerr nonlinearity saturates the photon number and the effective chemical potential is large and positive. High-energetic states are populated while states close to $\ketbra{0}_\ell$ are unoccupied, as one can notice from the annular shape of $W_\ell(\alpha, \alpha^*)$.
    This is the typical signature of population inversion.
    Throughout the prethermal domain, local states accumulate entropy, as it results from the monotonic growth of $\delta n_\ell$, $T_\ell$ and $S_\ell$.
    \item [(iii)] \textit{$\mathbb{U}(1)$-symmetric thermal domain}. The Gibbs state \eqref{eqs:gibbs} in the limit $U=0$ describes the local physics of the system. The Kerr nonlinearity does play a marginal role in shaping $\hat{\rho}_\ell$, which reduces to the vacuum $\ketbra{0}_\ell$ dressed by linear thermal fluctuations. 
    The temperature can be extracted from the equipartition theorem and 
    the chemical potential $\mu_\ell$ is negative.
    The local states rapidly lose entropy while approaching the undriven edge of the chain.
\end{enumerate}

\begin{figure*}[t!]
\centering
\includegraphics[width=0.85 \textwidth]{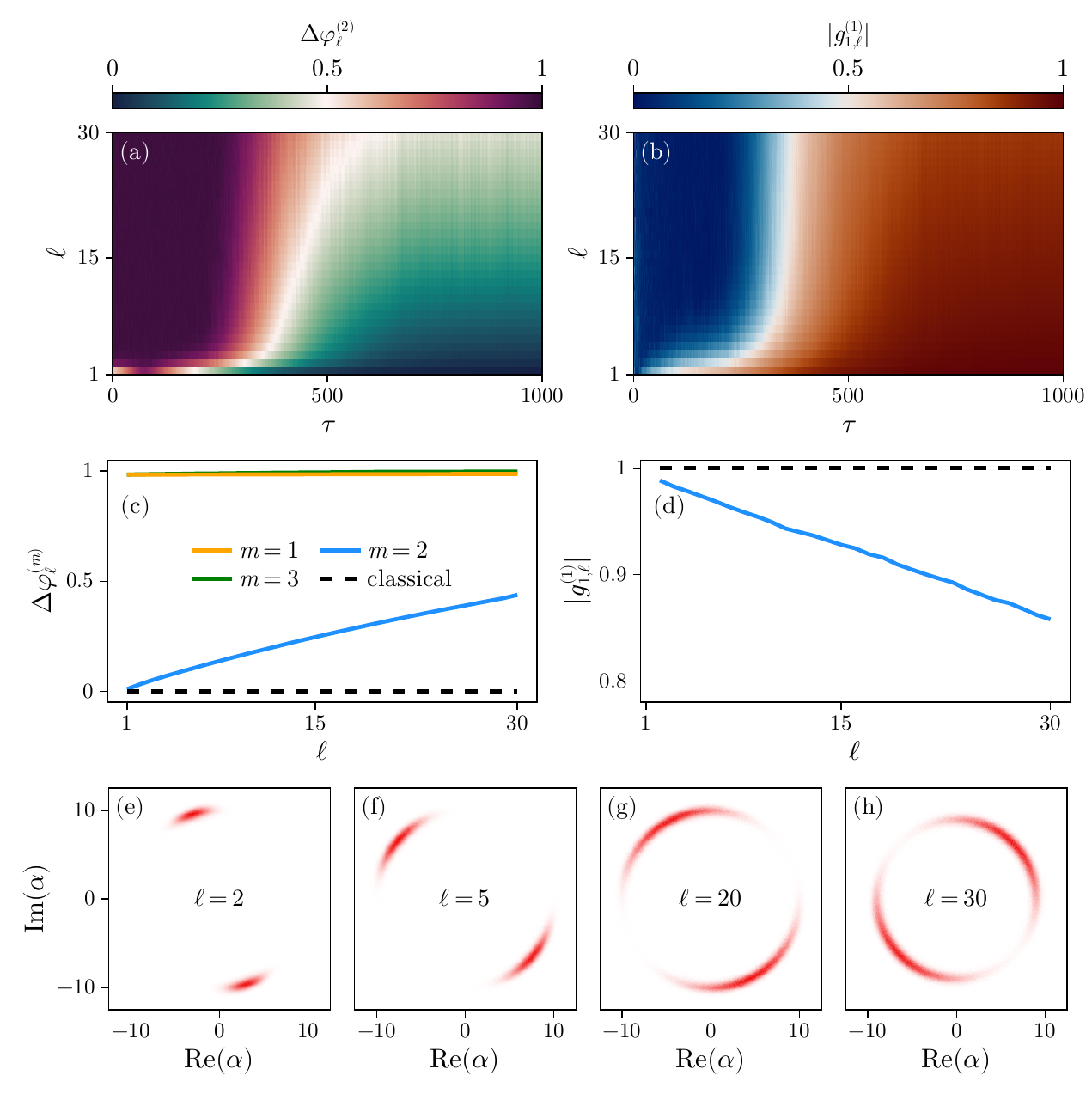}\vspace{0.7em}
\caption{
Regular RNW regime for $n=2$.
(a) Dynamics of the phase variance $\Delta\varphi_\ell^{(2)}$ as a function of time $t$ and site index $\ell$ in a $L=30$ chain.
(b) same as in panel (a) but for the first-order coherence function $|g_{1,\ell}^{(1)}|$.
(c) Steady-state profile of $\Delta\varphi_\ell^{(m)}$ for $m=1, 2, 3$ (orange, blue and green lines respectively).
The black-dashed line corresponds to $\Delta\varphi_\ell^{(1)}$ computed by solving the classical Gross-Pitaevskii equations of motion.
(d) Steady-state profile of $|g_{1,\ell}^{(1)}|$. The black-dashed line indicates the Gross-Piatevskii solution.
(e-h) Local Wigner functions $W_\ell(\alpha, \alpha^*)$ for (e) $\ell=2$, (f) $\ell=5$, (g) $\ell=20$, and (h) $\ell=30$. 
The drive amplitude is set to $\zeta=3.5$.
The other parameters are set as in Fig.~\ref{fig:phase_diagram}.
}
\label{fig:rnw_2}
\end{figure*}

\begin{figure*}[t!]
\centering
\includegraphics[width=0.95 \textwidth]{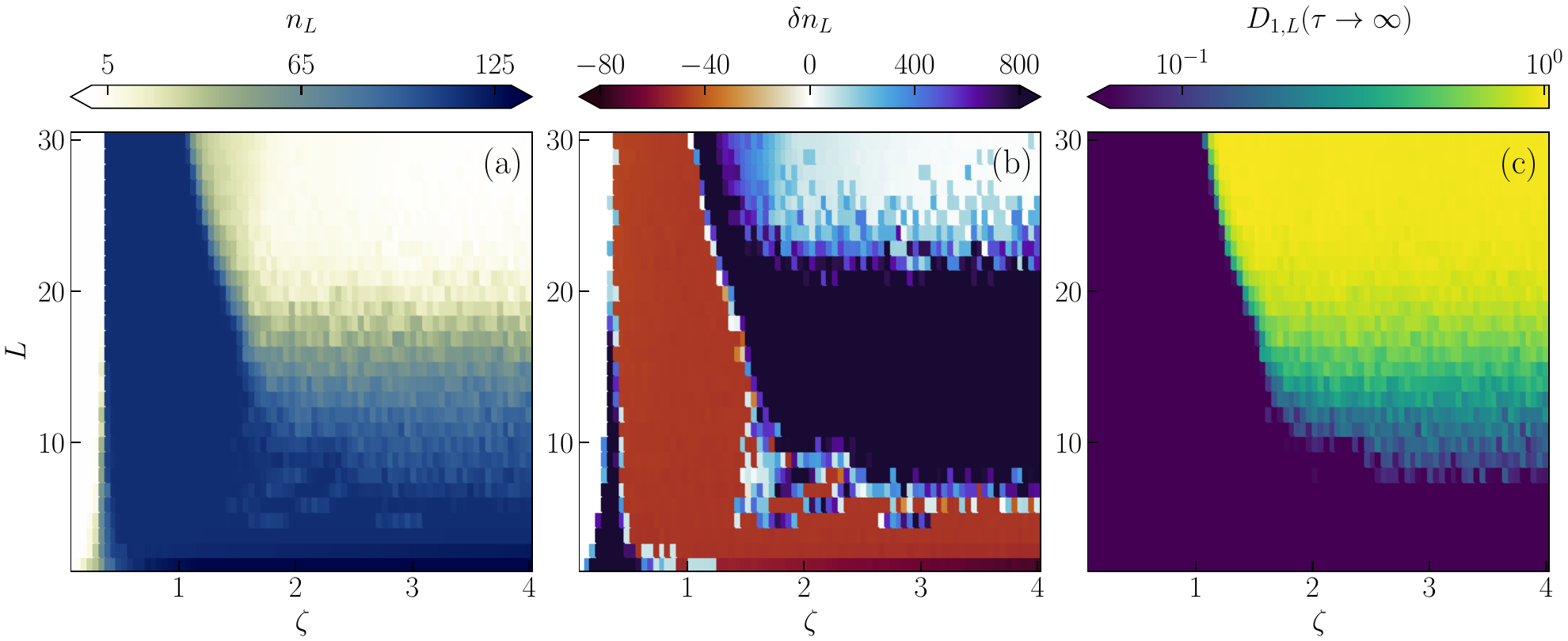}\vspace{0.8em}
\caption{
Phase diagram in the NESS as a function of the chain length $L$ and the 3-photon drive amplitude $\zeta$ for the last resonator $\ell=L$.
(a) Photon number $n_L$.
(b) Photon fluctuations $\delta n_L$, computed according to Eq.~\eqref{eqs:photon_fluctuations}.
(c) Saturation value of the semiclassical OTOC $D_{1,L}(\tau\to\infty)$, computed according to Eq.~\eqref{eqs:semiclassical_OTOC}.
Results for panels (a) and (b) are computed by averaging over $N_{\rm traj} = 10^3$ independent trajectories, while results in panel (c) over $10^2$ trajectories. 
Statistics are further improved by averaging over a time window $\Delta \tau = 10^3$ after reaching the steady state. 
In this section, we fix $\Delta=7$ and $J=4$.
The other parameters are set as in Fig.~\ref{fig:phase_diagram}.
}
\label{fig:phase_diagram_3}
\end{figure*}

This counterintuitive behavior of $T_\ell$, referred as to anomalous heating, in the prethermal domain has been observed in Ref.~\cite{ferrari_chaos_2024} and in discrete nonlinear Schr\"odinger equations coupled to thermal baths at the boundaries \cite{iubini_nonequilibrium_2012}.
It originates from the coupling between particle and heat transport.
In the bulk of the chain, the transport of particles dominates over the transport of heat, due to the large and positive chemical potential, which strives to add particles to the local sites, whose population is saturated by the finite Kerr nonlinearity.
At the end of the chain, $\mu_\ell$ becomes negative and the conventional heat flow dominates over the photon transport, leading to a negative thermal gradient.

Some final comments are in order.
In the hydrodynamic regime, the symmetry of the boundary drive influences only the first few sites, over a microscopic length scale. 
The restoration of the $\mathbb{U}(1)$ symmetry leads to the very same regimes observed in the chaotic regime of Ref.~\cite{ferrari_chaos_2024}, indicating the universality of the prethermal and thermal domains in a broad class of boundary-driven, dissipative nonlinear bosonic chains. 
Here, we employed a Gibbs ansatz, which naturally deals with the on-site Kerr nonlinearity. 
In the Appendix \ref{sec:appendix_C}, we show that the prethermal and thermal domain can also be described by a $\mathbb{U}(1)$-symmetric driven-dissipative impurity ansatz, where the nonlinearity is encoded in the Lindblad jump operators.

\subsection{Regular RNW regime}\label{sec:nrw_regime}

\begin{figure*}[t!]
\centering
\includegraphics[width=0.9 \textwidth]{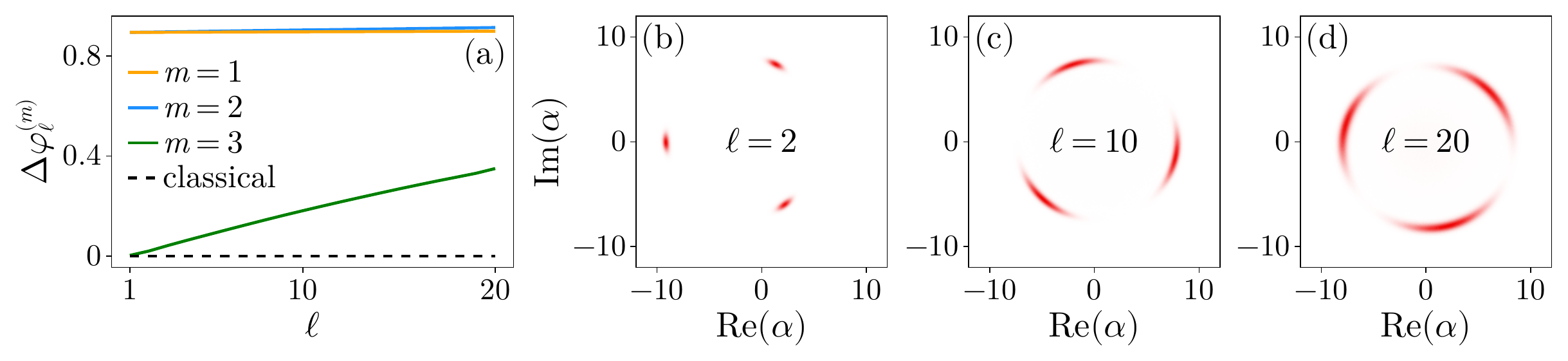}\vspace{0.8em}
\caption{
Regular RNW regime for $n=3$. 
(a) Steady-state profile of the circular variances $\Delta\varphi_\ell^{(m)}$ for $m=1, 2, 3$ (orange, blue and green line respectively). The black-dashed line corresponds to $\Delta\varphi_\ell^{(1)}$ computed solving the classical Gross-Pitaevskii equations of motion.
(b-d) Local Wigner functions $W_\ell(\alpha, \alpha^*)$ for (b) $\ell=2$, (c) $\ell=10$, and (d) $\ell=20$.
The drive amplitude is set to $\zeta=1.4$.
The other parameters are set as in Fig.~\ref{fig:phase_diagram_3}.
}
\label{fig:rnw_3}
\end{figure*}

In this section we focus on the sub-Poissonian resonant nonlinear wave (RNW) regime, which shows remarkable differences with respect to the hydrodynamic regime described in the previous section: (i) distinct quantum features appear, (ii) the symmetry of the boundary drive is trasferred to the whole chain. 
Throughout the section, we fix the 2-photon drive amplitude to $\zeta=3.5$.

We first investigate how quantum fluctuations impact the dynamics of the RNW regime. 
We introduce the $m$-th circular moment $C^{(m)}_\ell = |\langle \rme^{\rmi m \varphi_\ell}\rangle|$, with $m\in\mathbb{N}$, and define the circular variance of order $m$, $\Delta\varphi_\ell^{(m)} = 1 - C_\ell^{(m)}$.
Given a $m$-modal probability distribution in the phase space, which in the limiting case consists simply of $m$ symmetrically organized points on a circle with fixed radius, only the $m$-th circular momentum $C^{(m)}$ is nonzero.
A decrease of $C^{(m)}$ (or, equivalently, an increase of $\Delta\varphi^{(m)}$) signals phase spreading on the circle.
In the case of uniformly distributed phases, circular momenta vanish for any $m$.
In this section, we monitor $\Delta\varphi_\ell^{(m)}$ for different orders $m$ as a function of the site index $\ell$.

We also consider the first-order coherence function defined as
\begin{equation}
    g_{k,\ell}^{(1)} = \frac{\langle\hat{a}_k^\dagger\hat{a}_\ell\rangle}{\sqrt{\langle\hat{a}_k^\dagger\hat{a}_k\rangle\langle\hat{a}_\ell^\dagger\hat{a}_\ell\rangle}},
\end{equation}
which interpolates from disordered systems where $|g_{k,\ell}^{(1)}|\sim \rme^{-\Lambda|k-\ell|}$, being $\Lambda$ a microscopic correlation length, to long-range ordered systems where $|g_{k,\ell}^{(1)}|\sim \textit{const}$. 
In Figs.~\ref{fig:rnw_2} (a) and (b) we plot the spatiotemporal evolution of $\Delta\varphi_\ell^{(2)}$ and $|g_{1,\ell}^{(1)}|$ for a $L=30$ chain in the RNW regime.
Both quantities show an initial transient where $\Delta\varphi_\ell^{(2)}\simeq1$ and $|g_{1,\ell}^{(1)}|\simeq0$ in the bulk, followed by a steady state where phase correlations spread over the whole chain and the circular variance is smaller than 1, signaling that the steady state is no longer $\mathbb{U}(1)$ invariant.
In Figs.~\ref{fig:rnw_2} (c) and (d) we plot the profile of $\Delta\varphi_\ell^{(m)}$ for $m=1, 2, 3$ and $|g_{1,\ell}^{(1)}|$ in the steady state.
We also consider the classical Gross-Pitaevskii solution (black-dashed line) for $\Delta\varphi_\ell^{(1)}$ and $|g_{1,\ell}^{(1)}|$.
Quantum mechanically, $\Delta\varphi_\ell^{(m)}$ for $m=1$ and $3$ stays close to one, indicating that the circular moments are not resonant with $W_\ell(\alpha, \alpha^*)$ in the phase space. 
Conversely, $\Delta\varphi_\ell^{(2)}\simeq 0$ for small $\ell$, proving that $W_\ell(\alpha, \alpha^*)$ is a well-defined bimodal distribution.
We observe that $\Delta\varphi_\ell^{(2)}$ increases across the chain, signaling a phase spreading away from the driven edge.
The classical equations of motion, instead, predict $\Delta\varphi_\ell^{(1)}=0$ regardless of $\ell$.
We conclude that the growth of phase uncertainty in the RNW regime is due to quantum fluctuations.
Similarly, quantum fluctuations downgrade the classical long-range order to quasi-long range order in the steady state, inducing the decay of $g^{(1)}_{1, \ell}$. 

Notably, the boundary 2-photon drive shapes the local states throughout the entire chain. 
In Figs.~\ref{fig:rnw_2} (a-d) we plot the local Wigner function for $\ell=2$, $\ell=5$, $\ell=20$ and $\ell=30$.
The $\mathbb{Z}_2$ symmetry of the drive remains imprinted across the chain, and $W_\ell(\alpha, \alpha^*)$ draws a couple of arcs in phase space with an increasing angular spreading, encoded by $\Delta\varphi_\ell^{(2)}$, and a radial localization, underpinning the sub-Poissonian statistics.

\section{3-photon driven bosonic chain}\label{sec:3_photon}

In this section, we consider boundary higher-photon drives, by fixing $n=3$ in Eq.~\eqref{eqs:hamiltonian}. We consider the following parameters as fixed: $\Delta=7$, $J=4$.
We find that the qualitative phase diagram of the system does not change for other choices of these parameters.

First, we explore the phase diagram of the system by monitoring $n_\ell$, $\delta n_\ell$ and $D_{1, L}(\tau\to\infty)$ as $L$ and $\zeta$ are varied \footnote{For $n>2$ the vacuum becomes semiclassically stable \cite{minganti_dissipative_2023}, significantly slowing down the typical times needed to reach the steady state. To speed up this timescale, and thus reduce the computational cost of evolving stochastic trajectories for very long times, we choose the initial state for Eqs.~\eqref{eqs:stochastic_differential_equations} as $\alpha_\ell(0) = r[\cos(\theta) + \rmi\sin(\theta)] + \xi_\ell$ with $\theta$ uniformly distributed within $[-\pi, \pi]$, $\xi_\ell\sim\mathcal{N}(0, 1/2)$, and $r=10$.
We consider these initial conditions for all simulations with $n>2$.}.
Results are collected in Fig.~\ref{fig:phase_diagram_3}.
The structure of the 3-photon drive phase diagram resembles the one of the 2-photon drive phase diagram in Fig.~\ref{fig:phase_diagram}. 
In Fig.~\ref{fig:phase_diagram_3} we distinguish: (I) a regular quasilinear regime characterized by Poissonian fluctuations and low photonic population, as well as by a vanishing semiclassical steady-state OTOC; (II) the RNW regime, characterized by sub-Poissonian statistics, $D_{1, L}(\tau\to\infty) \simeq 0$, and a large photonic population; (III) a hydrodynamic regime, where super-Poissonian statistics emerges and the OTOC signals chaotic behavior; (IV) a thermal vacuum regime, where $\delta n_L$ approaches the Poissonian limit and $D_{1, L}(\tau\to\infty) \simeq 1$.

To characterize the RNW regime,
we consider again the $m$-th order circular variance $\Delta\varphi_\ell^{(m)}$ for $m=1, 2, 3$.
In Fig.~\ref{fig:rnw_3} (a) we plot $\Delta\varphi_\ell^{(m)}$ against $\ell$ for a $L=20$ chain.
For $m=1, 2$, $\Delta\varphi_\ell^{(m)}\simeq1$, as the circular moments are not resonant with $W_\ell(\alpha, \alpha^*)$.
$\Delta\varphi_\ell^{(3)}$ grows monotonically across the chain, staying close to zero for small $\ell$.
This behavior encodes the $n$-modality of $W_\ell(\alpha, \alpha^*)$ in the RNW regime and the increasing phase decoherence with $\ell$.
The black-dashed line represents $\Delta\varphi_\ell^{(1)}$ computed from the Gross-Piatevskii solution.
The vanishing classical circular variance proves the quantum origin of the phase spreading detected by the TWA.

In Figs.~\ref{fig:rnw_3} (b-d), we plot the local Wigner function for $\ell=2$, $\ell=10$, and $\ell=20$.
The $\mathbb{Z}_3$ symmetry of the coherent drive is imprinted throughout the driven-dissipative chain.
The Wigner function consists of three arcs in phase space, each one characterized by a growing phase variance induced by quantum fluctuations.
As before, the radial localization of $W_\ell(\alpha, \alpha^*)$ induces the sub-Poissonian statistics of the bosonic field.

\section{Conclusion}\label{sec:conclusion}

We have studied the dynamical properties and the NESS of boundary-dissipative nonlinear bosonic lattices in one dimension, subject to a $n$-photon drive on the first site.
While the NESS phase diagram has been previously explored for single-photon driving \cite{prem_dynamics_2023, kumar_observation_2024, ferrari_chaos_2024}, our results identify which dynamical regimes remain sensitive to the symmetry and structure of the boundary drive.
This analysis advances the understanding of quantum state engineering in open many-body systems and suggests concrete strategies for realizing nonclassical, correlated steady states in superconducting circuit QED platforms.

Our results reveal that the sensitivity of the bulk NESS to the boundary conditions is tightly linked to the presence or absence of chaotic dynamics.
In the chaotic regime, the $\mathbb{Z}_n$ symmetry imposed by the drive is rapidly erased by long-wavelength hydrodynamic modes, restoring the global $\mathbb{U}(1)$ symmetry of the underlying Hamiltonian.
The system exhibits a two-stage thermalization process in space, leading to an extended ``prethermal" domain characterized by photon saturation, population inversion, and anomalous heating—phenomena analogous to those observed under single-photon driving in Ref.~\cite{ferrari_chaos_2024}.

In contrast, the regular, sub-Poissonian RNW regime exhibits pronounced sensitivity to the boundary drive.
The local quantum states along the entire chain retain the $\mathbb{Z}_n$ symmetry imposed at the boundary and the local driven-dissipative mechanisms determine the precise features of the full setup.
In addition, phase decoherence induced by quantum fluctuations broadens the local Wigner function along the circle, without affecting the radial localization underpinning the sub-Poissonian statistics.

A natural extension of this work involves incorporating $m$-photon dissipation at the two edges of the chain.
Such mechanisms could potentially stabilize genuinely nonclassical NESS, featuring negativity in the Wigner quasi-probability distribution and strong quantum correlations.
Capturing these effects theoretically would require methods beyond the truncated Wigner approximation, which cannot account for Wigner negativity \cite{carmichael_statistical_1999}.
Tensor network techniques--capable of simulating open quantum systems with large Hilbert spaces--represent a promising direction for addressing this challenge \cite{jaschke_one-dimensional_2018, weimer_simulation_2021}.

\begin{acknowledgments}
We acknowledge useful discussions with Catalin Halati.
This work was supported by the Swiss National Science Foundation through Project No. 200020\_215172.
\end{acknowledgments}

\appendix

\section{Details on methods}\label{sec:appendix_A}

\subsection{Derivation of the TWA}\label{sec:appendix_A1}

The truncated Wigner approximation relies on the phase-space representation of quantum mechanics.
Within this formalism, operators in the Hilbert space are mapped as phase space functions \cite{polkovnikov_phase_2010}.
The non-commutativity of quantum operators is encoded into an algebraic structure called the Moyal product, which can be operatively used to derive the phase-space version of quantum operators in a given basis, such as the coherent-state basis.
We refer the Reader to \cite{polkovnikov_phase_2010} for a comprehensive review of quantum mechanics in phase space.

The Lindblad equation \eqref{eqs:lindblad} is mapped in a partial differential equation (PDE) for a multidimensional, time-dependent phase-space function called Wigner function $W(t; \alpha_1, \alpha_1^*,\, ...,\,\alpha_L, \alpha_L^*)$. The Wigner function is the phase-space representation of the many-body density matrix. The PDE for $W$ reads
\begin{widetext}
\begin{align}\label{eqs:PDE}
   \rmi \frac{\partial W}{\partial t} = & - \Delta\sum_{\ell=1}^L\left(\alpha_\ell^*\frac{\partial}{\partial\alpha_\ell^*} - \alpha_\ell\frac{\partial}{\partial\alpha_\ell}\right)W 
    - J\sum_{\ell=1}^{L-1}\left(\alpha_{\ell+1}^*\frac{\partial}{\partial\alpha_\ell^*} - \alpha_{\ell+1}\frac{\partial}{\partial\alpha_\ell}\right)W
    \\& - \frac{\zeta}{n}\sum_{k=0}^{+\infty}\frac{n!\,2^{-2k}}{(2k+1)!(n-2k-1)!}\left[\alpha_1^{n-2k-1}\frac{\partial^{2k+1}}{\partial\alpha_1^{* 2k+1}} - (\alpha_1^*)^{n-2k-1}\frac{\partial^{2k+1}}{\partial\alpha_1^{2k+1}}\right]W\nonumber
   \\
   & 
   + U\sum_{\ell=1}^L(|\alpha_\ell|^2-1)\left(\alpha_\ell^*\frac{\partial}{\partial\alpha_\ell^*} - \alpha_\ell\frac{\partial}{\partial\alpha_\ell}\right)W - \frac{U}{4}\sum_{\ell=1}^L\left(\alpha_\ell^*\frac{\partial}{\partial\alpha_\ell^*} - \alpha_\ell\frac{\partial}{\partial\alpha_\ell}\right)\frac{\partial^2}{\partial\alpha_\ell\partial\alpha_\ell^*}W \nonumber \\
&   + \frac{\rmi\gamma}{2}  \left[ 
\frac{\partial}{\partial\alpha_1}(\alpha_1 W)+ \frac{\partial}{\partial\alpha_1^*}(\alpha_1^*W) + \frac{\partial^2}{\partial\alpha_1\partial\alpha_1^*}W
+ \frac{\partial}{\partial\alpha_L}(\alpha_L W)+ \frac{\partial}{\partial\alpha_L^*}(\alpha_L^*W) + \frac{\partial^2}{\partial\alpha_L\partial\alpha_L^*}W
\right].\nonumber
\end{align}
\end{widetext}
The truncated Wigner approximation corresponds to an expansion of the Moyal product up to second order \cite{ferrari_chaos_2024}.
This allows the mapping of the Lindblad equation into a Fokker-Planck equation for the now well-defined probability distribution $W$.
Equivalently, this amounts in neglecting the terms in Eq.~\eqref{eqs:PDE} involving derivatives of order higher than two, coming from the Kerr nonlinearity and from the multiphoton drive.
The resulting Fokker-Planck equation can be further mapped into the set of stochastic differential equations~\eqref{eqs:stochastic_differential_equations} (see Ref.~\cite{carmichael_statistical_1999} for an in depth discussion about the relation between Fokker-Planck and Langevin equations).
Notably, while the Kerr nonlinearity is assumed to be small throughout the paper, the drive amplitude is not.
In particular, drives with $n\ge3$ generate a finite high-order derivative which accounts for non-Gaussian quantum effects, whose strength is controlled by $\zeta$.
We therefore expect the TWA to be less precise, and eventually to break down, for strongly-driven chains with relatively large $n$.

Expectation values in the Hilbert space becomes statistical expectation values in the phase space, weighted by the Wigner function \cite{polkovnikov_phase_2010}.
Within the TWA, expectation values are computed by averaging the phase-space representations of operators over many stochastic solutions of Eqs.~\eqref{eqs:stochastic_differential_equations}.
Photon number and  correlations become
\begin{align}
    &\langle\hat{a}_\ell^\dagger\hat{a}_k\rangle = \langle \alpha_\ell^*\alpha_k\rangle - \delta_{\ell,k}/2,\\
    &\langle\hat{a}_\ell^{\dagger 2}\hat{a}_\ell^2\rangle = \langle|\alpha_\ell|^4\rangle - 2\langle|\alpha_\ell|^2\rangle + 1/2,\nonumber
\end{align}
where $\langle\,...\,\rangle$ denotes the statistical average over many independent stochastic trajectories solutions of Eqs.~\eqref{eqs:stochastic_differential_equations}.
The local Wigner function, representing the local density matrix $\hat{\rho}_\ell(t) = \operatorname{Tr}_{k\ne\ell}\left[\hat{\rho}(t)\right]$ can be reconstructed by realizing an histogram of the fields $\alpha_\ell(t)$.
Throughout the paper, we consider only the steady-state local Wigner function and we compute it by histogramming the data of the bosonic fields obtained from the time evolution of $N_{\rm traj}=225$ trajectories and from a time window of $\Delta\tau=10^4$ once the steady state is reached.

As initial condition for Eqs.~\eqref{eqs:stochastic_differential_equations}, unless otherwise stated, we choose the vacuum $\ketbra{0}_\ell$ by sampling $\alpha_\ell(0)$ from a Gaussian distribution with zero mean and variance 1/2. 
The numerical solution of Eqs.~\eqref{eqs:stochastic_differential_equations} has been obtained using the stochastic solver SOSRI within the package \textit{Stochastic Differential Equations} that is available in Julia. 

\subsection{Validity of the TWA}\label{sec:appendix_A2}

\begin{figure}[t!]
\centering
\includegraphics[width=0.47 \textwidth]{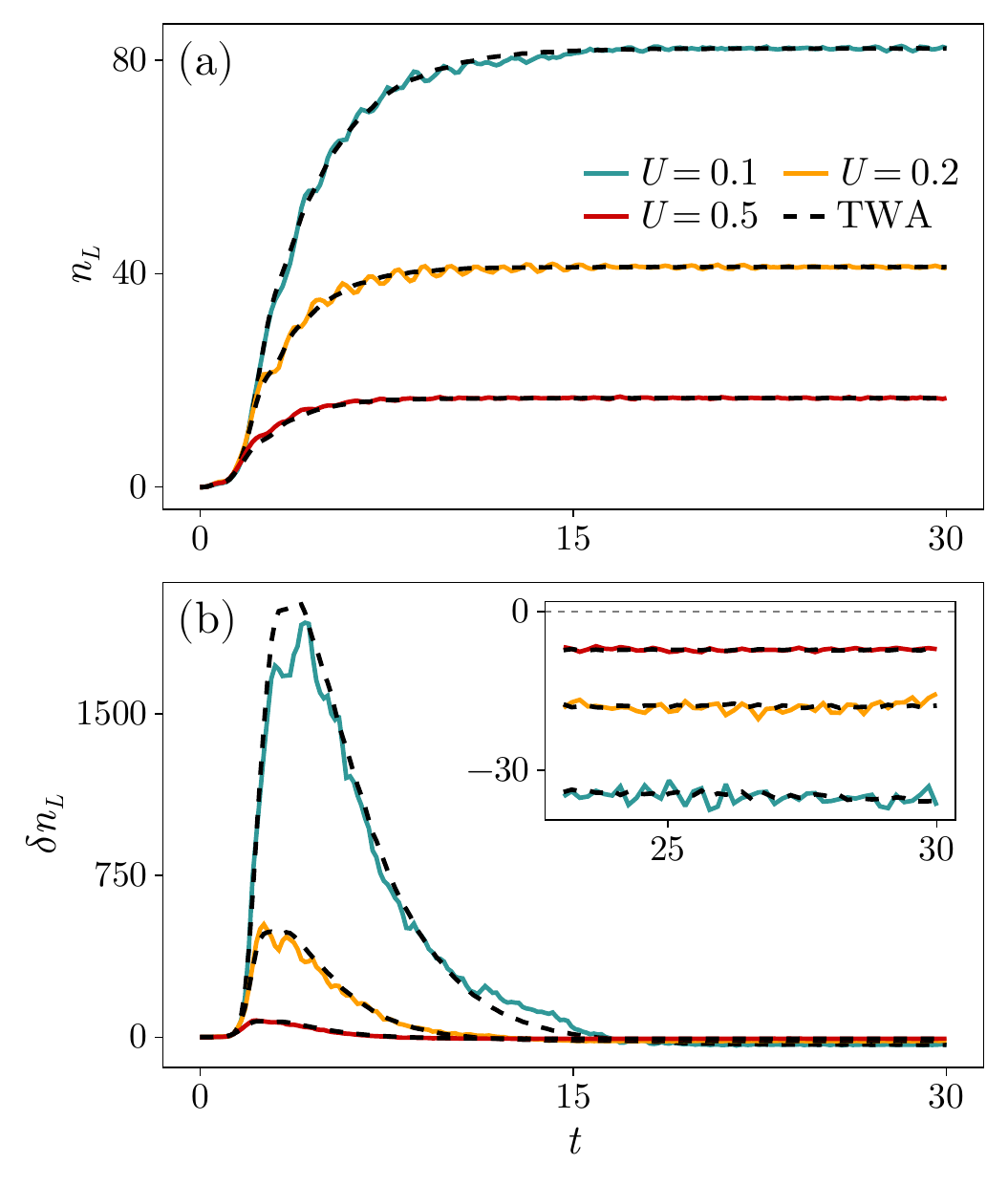}\vspace{0.1em}
\caption{
Comparison between TWA and exact dynamics for a $L=2$ chain subject to a boundary 2-photon drive.
(a) Time evolution of $n_L$ for $U=0.1$ (light blue curve), $U=0.2$ (yellow curve) and $U=0.5$ (red curve). The black-dashed line are the TWA data.
(b) As in (a) but for $\delta n_L$.
The inset in (b) shows the dynamics for times $t\ge20$.
TWA data are computed by averaging over $25\times 10^3$ stochastic trajectories, while for exact dynamics we averaged over $360$ quantum trajectories. 
The drive amplitude is fixed to $\zeta=4.5$. 
The other parameters are set as in Fig.~\ref{fig:phase_diagram}.
}
\label{fig:TWA_vs_ED}
\end{figure}

We benchmark the TWA against exact dynamics for the 2-photon driven chain for the smallest system size we considered, $L=2$.
We focus on the RNW regime and we compare the time evolution of $n_L$ and $\delta n_L$ for three different values of the nonlinearity, $U=0.1, 0.2, 0.5$.
Notice that the transformation $U\to U/r$, $r\in\mathbb{R}$ corresponds to the scaling towards the thermodynamic limit for which the classical equations of motion for $n=2$ remains unchanged (see the Appendix \ref{sec:appendix_A4} and Ref.~\cite{minganti_spectral_2018}).
Results are reported in Fig.~\ref{fig:TWA_vs_ED}.
Overall, we find a good agreement in both the chaotic, super-Poissonian transient and the regular, sub-Poissonian RNW steady state.
Both the photon number and its fluctuations match when computed with TWA and with exact dynamics.

Given the large photonic poulations for $U=0.1$ ($n_L\ge80$ in the steady state), to obtain exact quantities we averaged over many independent Monte Carlo quantum trajectories \cite{dalibard_wave-function_1992, molmer_monte_1993}, instead of time-evolving the density matrix with the Lindblad equation.
Quantum trajectories evolve the wave function $\ket{\psi(t)}$ in the Hilbert space, and the size of a single-cavity Hamiltonian is $d \times d$ where $d$ is the cutoff in the bosonic Hilbert space. 
The Lindblad equation evolves the density matrix $\hat{\rho}(t)$, and the size of a single-cavity Liouvillian is now $d^2 \times d^2$.
Quantum trajectories thus mitigate the computational overhead required by the Lindblad equation at the price of an average over many independent realizations.

Exact data in Fig.~\ref{fig:TWA_vs_ED} are obtained upon averaging over $360$ Monte Carlo quantum trajectories.
The numerical results for quantum trajectory dynamics is obtained with the \textit{QuantumToolbox.jl} package \cite{mercurio_ARXIV_2025} that is available in Julia.

\subsection{Semiclassical OTOC}\label{sec:appendix_A3}

\begin{figure}[t!]
\centering
\includegraphics[width=0.47 \textwidth]{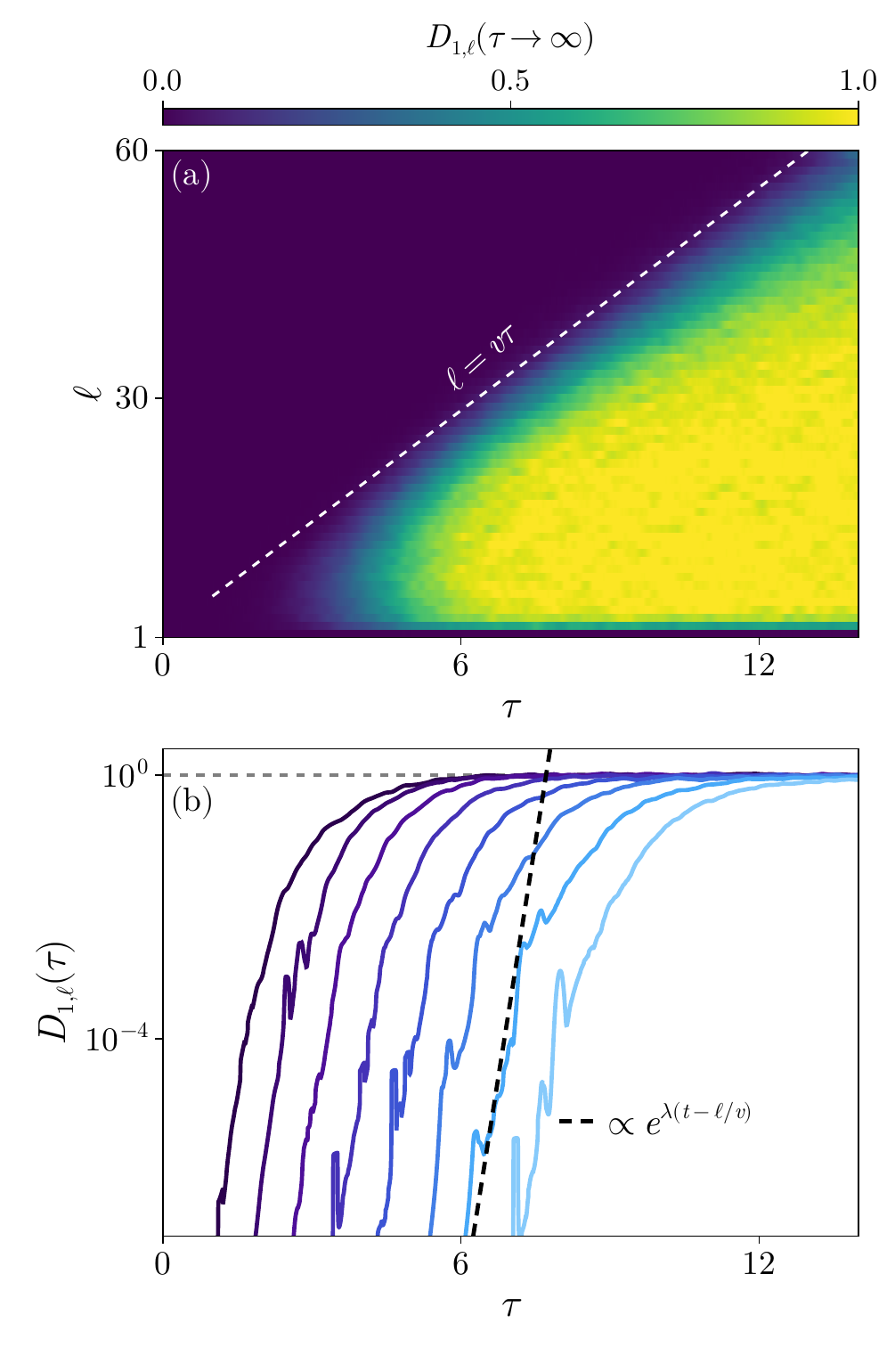}\vspace{0.1em}
\caption{
Spatio-temporal dynamics of the semiclassical OTOC defined in Eq.~\eqref{eqs:semiclassical_OTOC} in the 2-photon driven chain.
(a) Ballistic spreading of $D_{1, \ell}(\tau\to\infty)$ in a $L=60$ chain in the hydrodynamic regime.
The white-dashed line defines the butterfly velocity $v\simeq 2J$.
(b) Dynamics of $D_{1, \ell}(\tau\to\infty)$ as a function of time $\tau$ for $\ell=10,15,...,45$ (from dark to light blue).
The black-dashed line highlights the exponential growth of the semiclassical OTOC at a rate $\lambda\simeq 4.5$.
The data are computed by averaging over $10^3$ stochastic trajectories. 
The drive amplitude is fixed to $\zeta=10$. 
The other parameters are set as in Fig.~\ref{fig:phase_diagram}.
}
\label{fig:OTOC_dynamics}
\end{figure}

We derive here the expression of the semiclassical out-of-time-order correlator (OTOC). 
The quantum OTOC is defined as
\begin{align} \label{eq:SC-OTOC}
  C_{k,\ell}(t, \tau) \!:=\! \frac12  \textrm{Tr} \Big{(} [\hat B_\ell(t\!+\!\tau), \hat A_k(t)]^\dagger [\hat B_\ell(t\!+\!\tau), \hat A_k(t)]\hat \rho(t) \Big{)}.
\end{align}
Intuitively, $C_{k, \ell}$ measures how fast two local observables at sites $\ell$ and $k$ and times $t$ and $t+\tau$ stop commuting in the spacetime.
We decompose the annihilation operator as $\hat{a}_\ell = \sqrt{\hat{n}_\ell}\rme^{\rmi\hat{\varphi}_\ell}$ and we set $\hat{A}_\ell=\hat{n}_\ell$ and $\hat{B}_\ell=\rme^{\rmi\hat{\varphi}_\ell}$, satisfying $[\rme^{\rmi\hat{\varphi}_k}, \hat{n}_\ell] = \delta_{k\ell}\rme^{\rmi\hat{\varphi}_\ell}$.

Semiclassically, the quantum operators are replaced by $c$-numbers and the commutator becomes the Poisson bracket $\{n_k,\varphi_\ell\}=\delta_{k\ell}$ with
\begin{align}
    \{ f, g \} := \sum_{\ell=1}^L \left(\frac{\partial f}{\partial n_\ell} \frac{\partial g}{\partial \varphi_\ell}
    - 
    \frac{\partial g}{\partial n_\ell} \frac{\partial f}{\partial \varphi_\ell}\right)\,.
\end{align}
The semiclassical limit of Eq.~\eqref{eq:SC-OTOC} can be obtained with the above replacement and with the relation $\{ \varphi_\ell(t') , n_k(t) \} = - \delta \varphi_\ell(t') / \delta \varphi_k(t)$. It reads
\begin{align}\label{eqs:OTOC_definition2}
   D_{k,\ell}(t, \tau) & 
   = \frac{1}{2}\left\langle \left| \frac{\delta \rme^{\rmi \varphi_\ell(t+\tau)}}{\delta \varphi_{k}(t)} \right|^2\right\rangle \,,
\end{align}
where $\langle\,...\,\rangle$ now denotes the average over the quantum noise. 
The infinitesimal perturbation described by $\delta/\delta\varphi_\ell$ is implemented by realizing two copies $a$ and $b$ of the same system, differing by an infinitesimal phase perturbation at site $k$ and time $t$ in copy $b$.
The semiclassical OTOC then becomes
\begin{equation}
    D_{k, \ell}(t, \tau) = \frac{1}{2}\left\langle\left|\rme^{\rmi\varphi_\ell^{(a)}(t+\tau)} - \rme^{\rmi\varphi_\ell^{(b)}(t+\tau)}\right|^2\right\rangle\,,
\end{equation}
which leads to Eq.~\eqref{eqs:semiclassical_OTOC}.
The limit $D_{k,\ell}(\tau) = \lim_{t\to\infty}D_{k, \ell}(t, \tau)$ provides a diagnosis of steady-state chaos or integrability \cite{ferrari_dissipative_2025}, and throughout the paper we focus on $D_{k,\ell}(\tau)$ only.

Next, we analyze the spatiotemporal dynamics of $D_{k,\ell}(\tau)$ in the hydrodynamic regime studied in Sec.~\ref{sec:chaotic_regime}.
In Fig.~\ref{fig:OTOC_dynamics} (a) we plot the spatio-temporal evolution of $D_{1,\ell}(\tau\to\infty)$ in a $L=60$ chain for 
$\zeta=6$. 
The OTOC's dynamics describes the light cone of ballistic spreading of information typical of chaotic dynamics \cite{xu_scrambling_2024}. 
The butterfly velocity delimiting the light cone is the Lieb-Robinson velocity $v=2J$ \cite{lieb_finite_1972}.
In Fig.~\ref{fig:OTOC_dynamics} (b) we show the exponential growth in time of $D_{1,\ell}(\tau)\sim\rme^{\lambda(\tau - \ell/v)}$, where $\lambda\simeq 4.5$ is the Lypunov rate, a clear hallmark of chaotic semiclassical dynamics.

\subsection{Gross-Pitaevskii classical equations of motion}\label{sec:appendix_A4}

The classical limit of the Lindblad equation~\eqref{eqs:lindblad} can be obtained by replacing quantum operators with $c$-numbers.
This amounts to reducing quantum correlators into products of complex classical fields. From Eq.~\eqref{eqs:lindblad} one can derive the set of classical Gross-Pitaevskii equations of motion
\begin{align}\label{eqs:GP_equations}
    &\rmi \frac{\partial\alpha_1}{\partial t} = -\left(\Delta + \rmi \gamma/2\right)\alpha_1 + U|\alpha_1|^2\alpha_1\ - J\alpha_2 +F(\alpha_1^*)^{n-1}\,\nonumber,
    \\
    &\rmi \frac{\partial\alpha_\ell}{\partial t} = - \Delta\alpha_\ell + U|\alpha_\ell|^2\alpha_\ell - J \left(\alpha_{\ell+1} + \alpha_{\ell-1} \right)\,, \\
    &\rmi \frac{\partial\alpha_L}{\partial t} = -\left(\Delta + \rmi\gamma/2\right)\alpha_L + U|\alpha_L|^2\alpha_L - J\alpha_{L-1}\,,\nonumber
\end{align}
with $\ell=2,\,...\,L-1$. Eqs.~\eqref{eqs:GP_equations} time evolve the classical fields $\alpha_\ell(t)$ starting from $\alpha_\ell(0)=\alpha_0$ with $|\alpha_0|\ll1$.
Notably, Eqs.~\eqref{eqs:GP_equations} can be viewed as a zero-noise limit of the TWA equations~\eqref{eqs:stochastic_differential_equations} and the non-commutative Moyal product \cite{polkovnikov_phase_2010} reduces to the standard commutative product.

\section{Classical thermodynamics via the equipartition theorem}\label{sec:appendix_B}

\begin{figure}[t!]
\centering
\includegraphics[width=0.47 \textwidth]{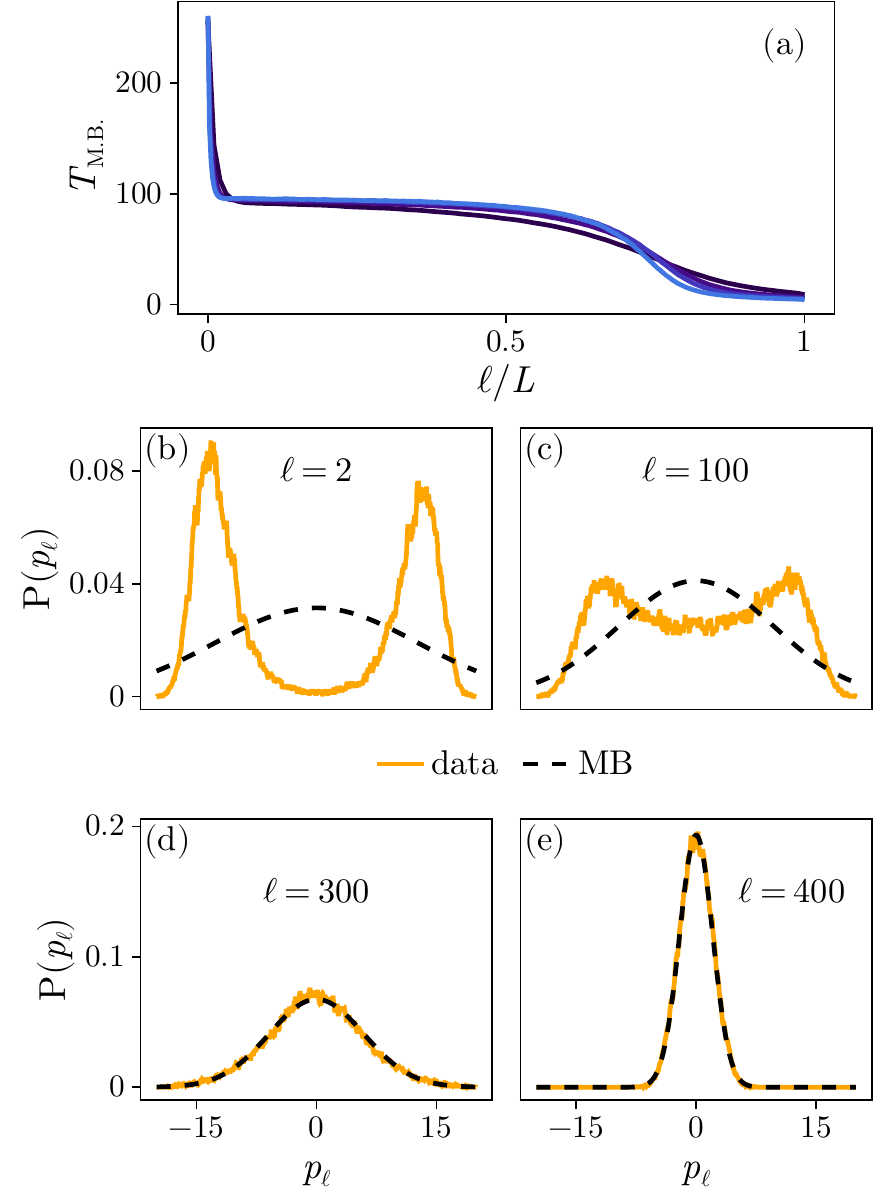}\vspace{0.1em}
\caption{
Application of the equipartition theorem in the chaotic 2-photon driven chain.
(a) Temperature $T_\ell$ estimated by applying the equipartition theorem as $T_\ell=|\Delta|\langle p^2_\ell\rangle$ for chains of size $L=100, 200, 300, 400$ (from dark to light blue).
(b-e) Comparison between $\textrm{P}(p_\ell)$ obtained from the TWA and the thermal Maxwell-Boltzmann distribution \eqref{eqs:MB} for (a) $\ell=2$, (b) $\ell=100$, (c) $\ell=300$, and (d) $\ell=400$ in a $L=400$ chain.
The drive amplitude is fixed to $\zeta=6$. 
The other parameters are set as in Fig.~\ref{fig:phase_diagram}.
}
\label{fig:equipartition}
\end{figure}

\begin{figure}[t!]
\centering
\includegraphics[width=0.47 \textwidth]{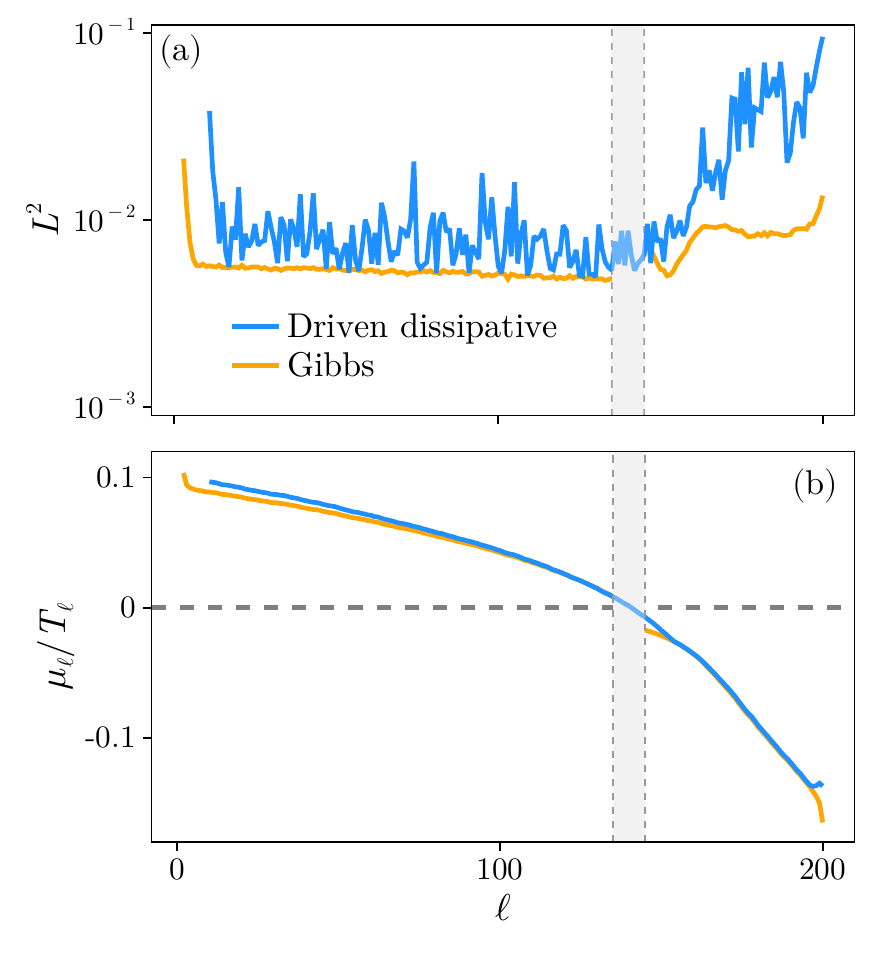}\vspace{0.1em}
\caption{
Comparison between the Gibbs and driven-dissipative ans\"atze for the local Wigner function $W_\ell(\alpha, \alpha^*)$ in a $L=200$ chain in the hydrodynamic regime.
(a) $L^2$ norms from the Gibbs ansatz \eqref{eqs:gibbs} [and the one-parameter ansatz \eqref{eqs:one_param}] (orange curve) and the driven-dissipative impurity model described by the jump operators \eqref{eqs:impurity_ansatz} (blue curve).
(b) Dimensionless quantity $\mu_\ell/T_\ell$ obtained from the two fitting models.
The drive amplitude is fixed to $\zeta=6$. 
The other parameters are set as in Fig.~\ref{fig:phase_diagram}.
}
\label{fig:L2}
\end{figure}

\begin{figure*}[t!]
\centering
\includegraphics[width=0.95 \textwidth]{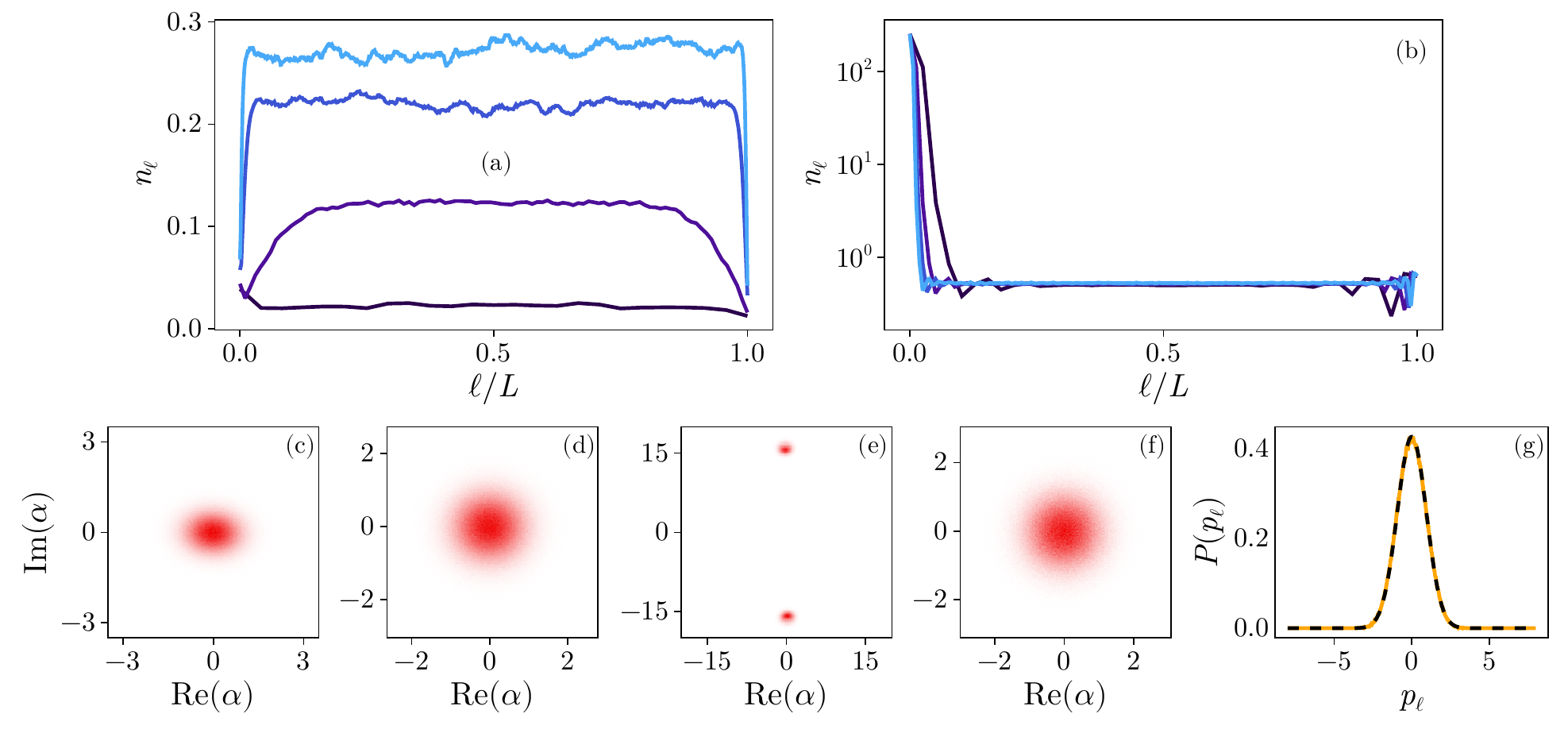}\vspace{0.8em}
\caption{
Additional regimes in the 2-photon driven, dissipative Bose-Hubbartd chain.
(a) Regular quasilinear regime. 
Photon number profile $n_\ell$ for chains with lengths $L=100, 200, 300, 400$ (from dark to light blue).
(b) and (c) Local Wigner functions $W_\ell(\alpha, \alpha^*)$ for $\ell=1$ and $\ell=L/2$ respectively in the $L=400$ chain.
The drive amplitude is fixed to $\zeta=1.5$.
(d) Thermal vacuum regime. 
Photon number profile $n_\ell$ for chains with lengths $L=100, 200, 300, 400$ (from dark to light blue).
(e) and (f) Local Wigner functions $W_\ell(\alpha, \alpha^*)$ for $\ell=1$ and $\ell=L/2$ respectively in the $L=400$ chain.
(g) Momentum distribution $\textrm{P}(p_\ell)$ for $\ell=L/2$, compared to the thermal Maxwell-Boltzmann distribution \eqref{eqs:MB}.
The drive amplitude is fixed to $\zeta=20$.
Results are computed by averaging over $N_{\rm traj} = 100$ independent trajectories. 
Statistics are further improved by averaging over a time window $\Delta \tau = 10^3$ after reaching the steady state. 
The other parameters are set as in Fig.~\ref{fig:phase_diagram}.
}
\label{fig:other_regimes}
\end{figure*}

In this section, we discuss whether the equipartition theorem can be applied to estimate the local effective temperature in the chain.
The equipartition theorem states that at thermal equilibrium, each quadratic degree of freedom contributes $T/2$ ($k_B=1$) to the local temperature \cite{toda_statistical_1983}.
It holds if quantum and nonlinear effects are negligible, and predicts that the local distribution of momenta follows a thermal Maxwell-Boltzmann distribution.
For a one-dimensional system like the one described by Eqs.~\eqref{eqs:hamiltonian} and \eqref{eqs:lindblad} one gets
\begin{equation}\label{eqs:MB}
    \textrm{P}_{\rm MB}(p_\ell) = \sqrt{\frac{|\Delta|}{2\pi T}}\rme^{-p^2/2T},
\end{equation}
where $|\Delta|$ fixes the energy scale of the quadratic part of Eq.~\eqref{eqs:hamiltonian}, and $T=|\Delta|\langle p^2 \rangle$ is the local temperature.
Within the TWA the momenta for the $\ell$-th site can be estimated as $\langle p_\ell \rangle=\sqrt{2}\langle\textrm{Im}(\alpha_\ell)\rangle$ and $\langle p_\ell^2 \rangle=2\langle\textrm{Im}^2(\alpha_\ell)\rangle$.

In Fig.~\ref{fig:equipartition} (a) we display the temperature profile estimated with the equipartition theorem for chains of increasing length in the hydrodynamic regime.
In the prethermal domain, the equipartition theorem predicts an almost flat, slowly decreasing $T_\ell$, in sharp contrast with the Gibbs temperature shown in Fig.~\ref{fig:hydrodynamics} (a), where anomalous heating was observed.
To explain this discrepancy, we compare in Figs.~\ref{fig:equipartition} (b-e) the momentum distributions $\textrm{P}(p_\ell)$ obtained from the histogram of $\langle p_\ell \rangle$ realized over $1.5\cdot 10^6$ total points in the steady state with the thermal Maxwell-Boltzmann distribution \eqref{eqs:MB}. 
The equipartition theorem breaks down over the entire prethermal phase due to the emergence of the Kerr nonlinearity that controls the saturation of the photon number.
At the end of the chain, instead, the local Wigner function is well captured by the thermal Maxwell-Boltzmann distribution, indicating that the Kerr term plays a marginal role and the local states are vacua dressed by linear thermal fluctuations.

\section{Driven-dissipative impurity ansatz for the hydrodynamic regime}\label{sec:appendix_C}

In the main text we fitted the local state $\hat{\rho}_\ell$ with an equilibrium Gibbs ansatz. 
We show here that $\hat{\rho}_\ell$ can also be reproduced by a local driven-dissipative impurity ansatz.
The single-site ansatz is given by the following Lindblad jump operators
\begin{align}\label{eqs:impurity_ansatz}
    &\hat{L}_\uparrow = \sqrt{\gamma^\uparrow} \hat a^\dagger,\qquad 
    \hat{L}_\downarrow = \sqrt{\gamma^\downarrow}\hat a,\qquad\\& 
    \hat{L}_\phi = \sqrt{\gamma^\phi}\hat{a}\hat a^{\dagger} ,\qquad 
    \hat{L}_{\rm s} = \sqrt{\gamma^{\rm s}} \hat a^2\,,\nonumber
\end{align}
where $\gamma_\uparrow$ and $\gamma_\downarrow$ control loss and incoherent gain mechanisms, $\gamma^{s}$ is a 2-photon decay, and $\gamma^\phi$ is a dephasing rate.
This model was introduced in Ref.~\cite{minganti_correspondance_2020} to study the connection of dissipative phase transitions and time crystals.
The system steady state describes a normal phase at $\gamma^{\downarrow}>\gamma^{\uparrow}$, characterized by vanishing photon number in the thermodynamic limit, and a saturated phase with a finite photon number at $\gamma^{\downarrow}<\gamma^{\uparrow}$.
The above model is valid if
\begin{equation}\label{eqs:SC_conditions}
    \gamma^{\uparrow} \sim \mathcal{O}(\gamma^{\downarrow}),\qquad \gamma^s\langle\hat{a}\hat{a}^{\dagger}\rangle\ll\gamma^{\uparrow}.
\end{equation}

One can associate a local temperature to the driven-dissipative impurity ansatz by interpreting $\hat{L}_\uparrow$ and $\hat{L}_\downarrow$ as couplings to linear thermal reservoirs. The detailed balance condition then sets the effective temperature as
\begin{equation}\label{eqs:detailed_balance}
    \frac{\gamma^\uparrow}{\gamma^\downarrow} = \rme^{- (\omega_0-\mu)/T} = \rme^{\mu/T},
\end{equation}
yielding $\mu/T = \log \left({\gamma^\uparrow}\big{/}{\gamma^\downarrow} \right)$,
where $\mu$ is an effective chemical potential.
Importantly, we can not extract independently $\mu$ and $T$ from the driven-dissipative ansatz, but only their ratio by fitting $\gamma^\uparrow$, $\gamma^\downarrow$ (and $\gamma^s$) on the local $W_\ell(\alpha, \alpha^*)$ throughout the chain.

First, we analyze the results of the 2D fit of $W_\ell(\alpha, \alpha^*)$ where the variational ansatz is either the Gibbs state in Eq.~\eqref{eqs:gibbs} or the driven-dissipative ansatz in Eq.~\eqref{eqs:impurity_ansatz}.
In Fig.~\ref{fig:L2} (a) we plot the $L^2$ norm between the system Wigner function and the fitted Wigner function, defined as
\begin{equation}\label{eqs:L2_norm}
    L^2(W_\ell, W^{\textrm{fit}}_\ell) = \left[\int_\mathcal{A}\rmd\alpha \rmd\alpha^* \, |W_\ell - W^{\textrm{fit}}_\ell|^2\right]^{1/2},
\end{equation}
for a $L=200$ chain in the hydrodynamic regime.
The orange curve represents the $L^2$ norm for the Gibbs state [or its $U=0$ limit in Eq.~\eqref{eqs:one_param}], while the blue curve represents the same quantity for the driven-dissipative ansatz.
Apart from the driven nonthermal domain, where neither of the two ans\"tze works, both the prethermal and thermal domains show that the two approaches reproduce the shape of $W_\ell(\alpha, \alpha^*)$.

In Fig.~\ref{fig:L2} (b) we plot the ratio $\mu_\ell/T_\ell$ for the same system size of Fig.~\ref{fig:L2} (a).
We observe that the profile of this dimensionless quantity is similar for both the Gibbs and the driven-dissipative ansatz.
Moreover, $\mu_\ell/T_\ell$ changes sign for both models at the same point, indicating the transition between the prethermal and thermal domains.
Similarly to Fig.~\ref{fig:hydrodynamics}, for the Gibbs fit we estimated $\mu_\ell/T_\ell$ with Eq.~\eqref{eqs:gibbs} in the prethermal domain and with Eq.~\eqref{eqs:one_param} in the thermal domain.
For the driven-dissipative impurity ansatz, $\mu_\ell/T_\ell>0$ implies $\gamma^\uparrow_\ell>\gamma^\downarrow_\ell$: in the prethermal domain, thermal gain dominates over photon losses, and the nonlinearity turns to be essential in preserving a finite site population.
We observe a population inversion, where high-energetic states are more populated than low-energy states.
On the contrary, $\mu_\ell/T_\ell<0$ implies $\gamma^\uparrow_\ell<\gamma^\downarrow_\ell$: here, incoherent losses dominates over saturation mechanisms, yielding a vacuum-like state dressed by thermal fluctuations.

\section{Additional regimes in the 2-photon driven chain}\label{sec:appendix_D}

In this section, we discuss the regimes (I) and (IV) that we identified in Fig.~\ref{fig:phase_diagram} and did not analyze in the main text.
In the regular quasilinear regime (I), the photon number acquires very low values across the full chain and independently of $L$, as shown in Fig.~\ref{fig:other_regimes} (a). 
The Wigner functions at $\ell=1$ [Fig.~\ref{fig:other_regimes} (b)] and $\ell=L/2$ [Fig.~\ref{fig:other_regimes} (c)] are slightly squeezed states.
In the thermal vacuum regime (IV), the photon number is large only in a microscopic sector of the chain, encompassing the very few first sites, as shown in Fig.~\ref{fig:other_regimes} (d).
In the bulk and at the undriven tail of the chain, $n_\ell$ drops below 1.
The Wigner function at $\ell=1$ [Fig.~\ref{fig:other_regimes} (e)] reflects the $\mathbb{Z}_2$ symmetry of the drive, while at $\ell=L/2$ [Fig.~\ref{fig:other_regimes} (f)] $W_\ell(\alpha, \alpha^*)$ displays a Gaussian shape.
In this case, $\textrm{P}(p_\ell)$ is well described by thermal Maxwell-Boltzmann distribution in agreement with the equipartition theorem, as shown in Fig.~\ref{fig:other_regimes} (g).
We point out that the truncated Wigner approximation may fail in describing passages of photonic clusters from the driven edge to the chain bulk in the thermal vacuum regime.
This could qualitatively change the observed photonic population profile, as well as the local states throughout the chain.
Capturing these possible nontrivial quantum effects may require different numerical techniques such as tensor network methods, and an in depth study goes beyond the scope of this work.

\providecommand{\noopsort}[1]{}\providecommand{\singleletter}[1]{#1}%

\end{document}